\documentclass[%
 aip,
 amsmath,amssymb,
 reprint,%
]{revtex4-1}

\usepackage{graphicx}%
\usepackage{dcolumn}%
\usepackage{bm}%

\usepackage[utf8]{inputenc}
\usepackage[T1]{fontenc}
\usepackage{mathptmx}
\usepackage{etoolbox}
\usepackage{xcolor}
\usepackage[normalem]{ulem}
\usepackage[version=4]{mhchem}
\usepackage{comment}
\usepackage{siunitx}
\usepackage{hyperref}
\newcommand{\silabel}{Supplementary Information}

\makeatletter
\def\@email#1#2{%
 \endgroup
 \patchcmd{\titleblock@produce}
  {\frontmatter@RRAPformat}
  {\frontmatter@RRAPformat{\produce@RRAP{*#1\href{mailto:#2}{#2}}}\frontmatter@RRAPformat}
  {}{}
}%
\makeatother
\begin{document}

\preprint{AIP/123-QED}

\title{High-pressure phase behaviors of titanium dioxide revealed by a $\Delta$-learning potential}
\author{Jacob G. Lee}
\affiliation{Department of Physics, Cavendish Laboratory, University of Cambridge, JJ Thompson Avenue, Cambridge, CB3 0HE}
 
\author{Chris J. Pickard}
\affiliation{Department of Materials Science \& Metallurgy, University of Cambridge, 27 Charles Babbage Road,
Cambridge, CB3 0FS, United Kingdom}
\affiliation{Advanced Institute for Materials Research, Tohoku University, Sendai, Japan}

 \author{Bingqing Cheng}%
 \email{bingqing.cheng@ist.ac.at}
\affiliation{The Institute of Science and Technology Austria, Am Campus 1, 3400 Klosterneuburg, Austria}%

\date{\today}%

\begin{abstract}

Titanium dioxide has been extensively studied in the rutile or anatase phases, while its high-pressure phases are less well understood, despite that many are thought to have interesting optical, mechanical and electrochemical properties.
First-principles methods such as density functional theory (DFT) are often used to compute the enthalpies of TiO$_2$ phases at 0~K, but they are expensive and thus impractical for long time-scale and large system-size simulations at finite temperatures.
On the other hand, cheap empirical potentials 
fail to capture the relative stablities of the various polymorphs.
To model the thermodynamic behaviors of ambient and high-pressure phases of TiO$_2$,
we design an empirical model as a baseline,
and then train a machine learning potential based on the difference between the DFT data and the empirical model.
This so-called $\Delta$-learning potential contains long-range electrostatic interactions,
and predicts the 0~K enthalpies of stable TiO$_2$ phases that are in good agreement with DFT.
We construct a pressure-temperature phase diagram of TiO$_2$ in the range $0<P<70$~GPa and $100<T<1500$~K.
We then simulate dynamic phase transition processes, by compressing anatase at different temperatures.
At 300~K, we observe predominantly anatase-to-baddeleyite transformation at about 20~GPa, via a martensitic two-step mechanism with highly ordered and collective atomic motion.
At 2000~K, anatase can transform into cotunnite around 45-55~GPa in a thermally-activated and probabilistic manner, accompanied by diffusive movement of oxygen atoms.
The pressures computed for these transitions show good agreement with experiments.
Our results shed light on how to synthesize and stabilize high-pressure TiO$_2$ phases,
and our method is generally applicable to other functional materials with multiple polymorphs.

\end{abstract}

\maketitle

\section{\label{sec:level1}Introduction} %

Titanium dioxide is widely used in a range of industries as a white pigment, and in biomedical applications such as drug delivery~\cite{Yin2013}.  As an intrinsic n-type semiconductor, it also has uses in solar cells~\cite{Mor2006}, photocatalysis~\cite{Hagfeldt1995}, and splitting water molecules for hydrogen production~\cite{Fujishima1972,Jeong2020}.
At ambient pressure, titanium dioxide has three common metastable crystalline polymorphs: anatase, rutile, and brookite, of which anatase and rutile are of most interest to industry.  There are also a large number of high-pressure metastable structures in the range 0-100~GPa, including TiO$_2$(B), hollandite, OI, fluorite, Pca2$_1$, ramsdellite, columbite (TiO$_2$(II)) and cotunnite~\cite{Dubrovinsky2001, AlKhatatbeh2009, Olsen1999}.  Many of these phases have interesting properties such as high optical absorbance~\cite{Dekura2011} and high hardness~\cite{Dubrovinsky2001, NishioHamane2010}. Crucially, many of these high-pressure structures remain stable at ambient pressures due to large kinetic barriers, hence it is useful to be able to understand the metastability of these phases and how they may be synthesised. 

Although many applications of ambient-pressure TiO$_2$ phases have been well-studied, 
the high-pressure phase behaviors are less understood.  Anatase, rutile, and brookite have been known since they can be found in mineral deposits.  Experimental work has shown the anatase-to-rutile and brookite-to-rutile transitions to be irreversible, suggesting rutile is the stable phase at low pressures~\cite{Hanaor2011, Liu2015, Gupta2011}.  The first high-pressure phase to be discovered was columbite~\cite{Dachille1968} which was made by compressing anatase and brookite using a diamond anvil and using x-ray powder diffraction to identify the structure.  Further experimental work using similar methods~\cite{Leger1993,Haines1993} demonstrated there is a series of high pressure polymorphs, and the pressures and temperatures at which transitions between these polymorphs occur have been determined~\cite{Sato1991, Olsen1999, AlKhatatbeh2009}.  More recent work has also suggested that phase transitions between polymorphs can be induced by shockwaves~\cite{Kalaiarasi2018} and lasers~\cite{Dauksta2019}.  
Nevertheless, the kinetic barriers of solid-solid transitions make precise experimental determination of the phase boundaries difficult.
Moreover, very little is known about the mechanisms of the these transitions, for example whether they are diffusive or martensitic.

A number of theoretical studies have used density functional theory (DFT) to determine the relative stability of the phases.  DFT generally shows very good agreement with experiment, except for the well-known problem that the enthalpy of rutile is significantly over-estimated, often suggesting anatase is the stable phase at 0~GPa~\cite{Muscat2002}, which is in disagreement with the experimental data cited above.  It is thought that this discrepancy may be due to unphysical delocalisation of titanium's 3d electrons~\cite{Luo2016}, but it is not well-understood.  Nevertheless, DFT, in particular using the PBEsol function, remains a useful tool for studying the high-pressure behavior~\cite{Mei2014, Samat2016, Dharmale2020}.  It is also possible to overcome the incorrect ordering of phases using semi-empirical methods, such as DFT+U~\cite{Arroyo2011} and DFT+D~\cite{Zhu2014}.  
Alternatively, Quantum Monte Carlo (QMC) methods including Variational Monte Carlo (VMC) and Finite Node Diffusion Monte Carlo (FN-DMC) require no empirical parameters, and have been shown to predict rutile as the 0~GPa stable phase~\cite{Luo2016} in line with experiment, however other studies have shown agreement with the DFT prediction~\cite{Needs2017}.

For problems that demand large system sizes and long simulation times, many studies utilise empirical potentials such as the Matsui-Akaogi (MA) potential~\cite{Matsui1991}, which has been shown to perform poorly compared to DFT, for example the incorrect prediction of the ordering of the stable phases~\cite{Reinhardt2019},
which contribute to the poor prediction of the TiO$_2$ pressure-temperature phase diagram compared to DFT or experiments~\cite{Reinhardt2020}.
Recently, some machine learning potentials of titanium dioxide have also been developed~\cite{Li2018}.

In the present work, we aim to understand the high-pressure phase behaviors of titanium dioxide, including the pressure-temperature phase diagram and the mechanism of solid-solid transitions.
To this end, we trained a machine learning potential (MLP) to accurately model all known and unknown phases of titanium dioxide for pressures up to 70~GPa and temperatures up to 2000~K based on DFT, but at orders of magnitude smaller computational cost.
Instead of learning the DFT energy directly, which is the usual strategy of fitting MLPs,
we first parameterise an empirical potential that includes long-range Coulomb interactions,
and then train a MLP to learn the difference between the DFT energy and the energy predicted by the empirical potential.
The combination of the MLP and the baseline empirical potential, which we call a $\Delta$-learning potential,
is inspired by the $\Delta$-machine learning approach that can promote the energies of small molecules computed by a cheap quantum mechanical method to a high-level of theory~\cite{Ramakrishnan2015}.
The $\Delta$-learning potential is then used to determine the relative metastability of the phases and investigate some of the phase transitions between anatase, baddeleyite, and cotunnite.

\begin{figure*}
    \centering
    \includegraphics[width=\linewidth]{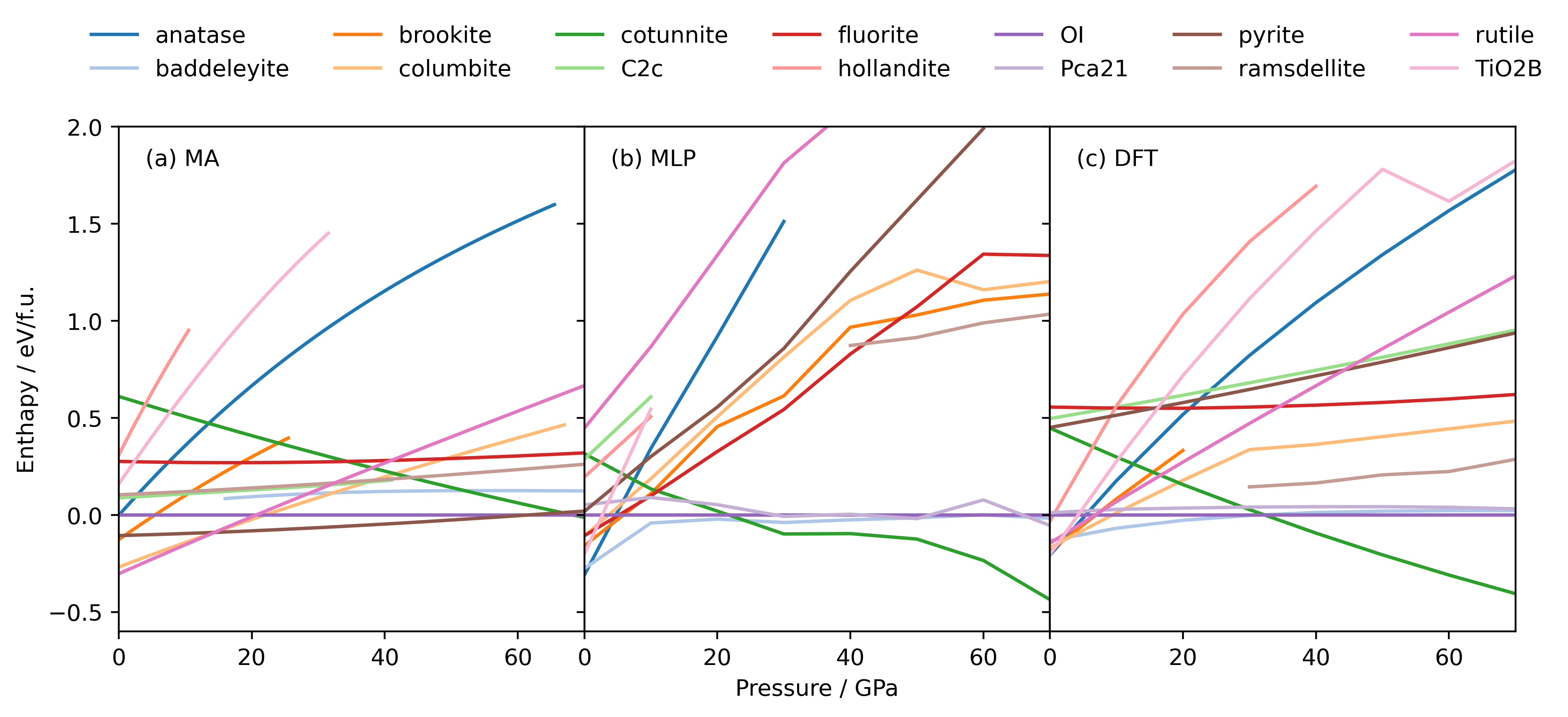}
    \caption{Enthalpies of known TiO$_2$ polymorphs at 0~K computed using (a) the MA potential, (b) the $\Delta$-learning MLP, and (c) PBEsol DFT. Note that for the MA potential the oxygen-displaced fluorite (P\text{ca}2$_1$) structure is not (meta)stable~\cite{Reinhardt2019}, and the ramsdellite and OI phases are somewhat different from the experimentally reported structures~\cite{Reinhardt2019}.}
    \label{fig:H}
\end{figure*}

\section{Methods}

\subsection{DFT reference}
We selected the PBEsol functional~\cite{perdew2008restoring} as it is designed to be most accurate in the high pressure limit and it generally shows better agreement with experimental lattice constants than the PBE and LDA functionals~\cite{Ding2014}.
It is worth mentioning that the accuracy of (semi-)local DFT functionals for TiO$_2$ at low pressure is subject to debate~\cite{lyle2015prediction}, since for example rutile is not predicted to be a stable phase,
although DMC shows good agreement in the ranking of static-lattice energies of phases with a number of DFT functionals~\cite{trail2017quantum}.
In addition, three functionals, LDA, PBE and PBEsol, all give consistent results regarding the ranking of stabilities for the known polymorphs of TiO$_2$~\cite{Reinhardt2020}.
We employed the CASTEP \textit{ab initio} simulation package~\cite{clark2005first}, and
full details of the DFT set-ups and configurations can be found in the input files supplied in the \silabel{}.
The 0~K enthalpies computed as above are shown in Fig.~\ref{fig:H}a.
The DFT data suggests that anatase should be the lowest enthalpy phase at very low temperature, followed by baddeleyite at 10 - 20~GPa, followed by cotunnite for $\geq$ 30~GPa.  Columbite is also competitive around 10~GPa, and OIphase and Pca2$_1$ are competitive around 30~GPa.  
Experimentally, it is well known that the stable phases under increasing compression follow the order rutile $\to$ columbite $\to$ baddeleyite $\to$ OI $\to$ cotunnite, which is true for HfO$_2$ and ZrO$_2$ as well as TiO$_2$~\cite{Dubrovinskaia2001}. Kinetic barriers make precise experimental determination of the phase diagram very challenging, however it is generally found that bulk anatase transforms to columbite at approximately 5~GPa.  In polycrystalline anatase, columbite is not formed, but baddeleyite is instead formed at 13~GPa~\cite{Arlt2000}.  On starting from bulk rutile, a transformation to columbite is sometimes seen at 7-10~GPa~\cite{Mammone1980,Arashi1992}, followed by a transformation to baddeleyite at 15~GPa, however, it is more common to see a transformation directly from rutile to baddeleyite near 12~GPa~\cite{Sato1991, Gerward1997}.  Baddeleyite transforms to cotunnite at approximately 60~GPa at 1700~K~\cite{Dubrovinsky2001}.
Hence, we may consider DFT data as being a good predictor for the purpose of determining the phase diagram, with the exception of the rutile phase as discussed above.

\subsection{Parameterising an empirical potential}

One of the most widely used empirical potentials for modelling titanium dioxide is the Matsui-Akaogi (MA) potential~\cite{Matsui1991}.  This includes both a Buckingham potential~\cite{Buckingham1938} and the usual Coulomb repulsion, where $q_1$ and $q_2$ are the effective charges of Ti and O:
\begin{equation}
    V = Ae^{-r/\rho} - \frac{C}{r^6} + \frac{q_1 q_2}{4\pi \varepsilon_0 r},
    \label{eq:MA}
\end{equation}
where $A$, $C$, and $\rho$ are empirical parameters.
However, when comparing the relative enthalpy of the metastable polymorphs computed by the MA potential with that computed by DFT, the performance of the MA potential is poor~\cite{Reinhardt2019}, as seen in Fig.~\ref{fig:H}.  In particular, from 15-65~GPa the MA potential predicts that pyrite is the stable phase, yet according to DFT pyrite has a significantly higher enthalpy than the stable phases, baddeleyite and cotunnite.  Despite this, it remains one of the most widely used potentials for the large time-scale and length-scale molecular dynamics simulations required to study nanoparticles~\cite{Naicker2005,Koparde2005}, surface physics~\cite{Heyhat2021}, crack propagation~\cite{Xu2016}, and other complex phenomena.

Here we adopt the functional form of the MA potential in Eqn.~\eqref{eq:MA}, and fitted the 
parameters $q_1$, $q_2$, $A$ and $C$. 
The fit was performed by minimizing the difference in the relative energies of 893 liquid TiO$_2$ configurations of 144 atoms compared to the PBEsol reference.
These liquid configurations were generated by molecular dynamics (MD) simulations employing the MA potential at $0-100$~GPa.
We call this new empircal potential for TiO$_2$ the reMA potential,
and the parameters are provided in the \silabel{}, as a LAMMPS input file.

\subsection{Training the MLP}

The MLP is constructed using the framework originally proposed by Behler and Parinello as implemented in RuNNer~\cite{Behler2007}.  90 symmetry functions are used, each modulated by a hyperbolic tangent cutoff function at up to 16$r_\text{Bohr}$.  The neural network consists of two hidden layers with 20 and 10 nodes respectively, and each with a hyperbolic tangent activation function.

\begin{figure}
    \centering
    \includegraphics[width=\linewidth]{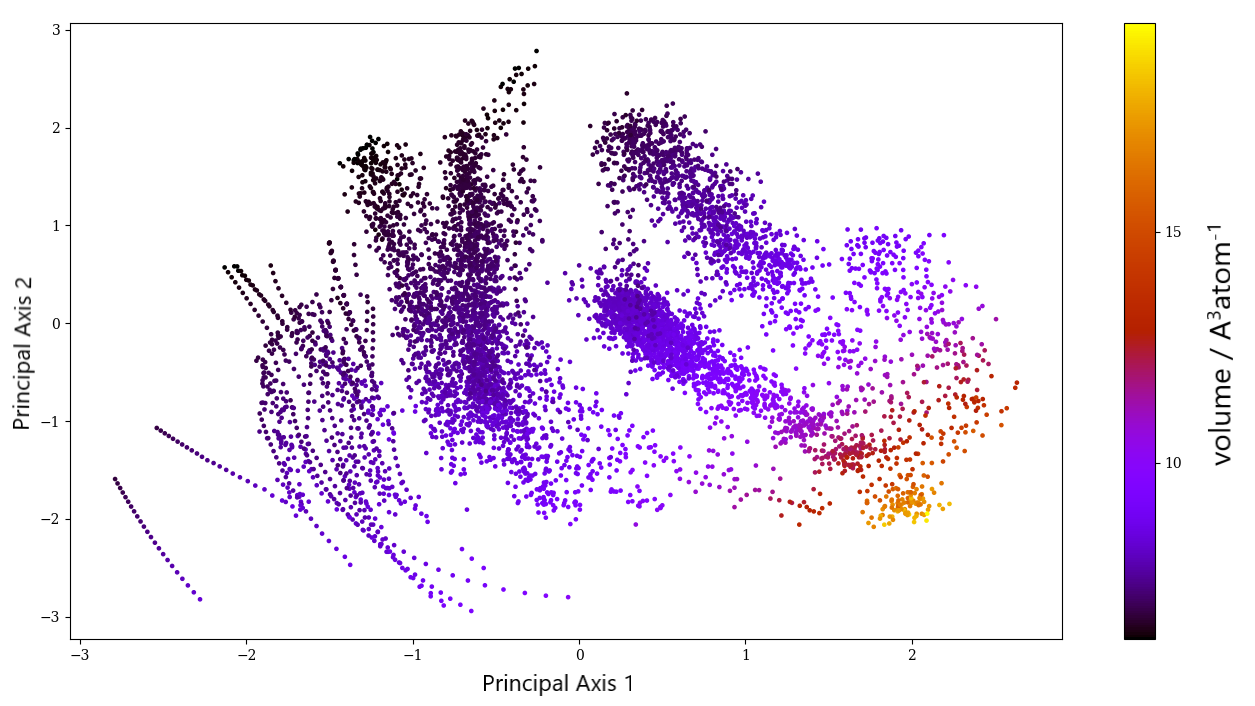}
    \caption{Projection of the training dataset onto its principal axes, generated using the ASAP package~\cite{Cheng2020asap}. }
    \label{fig:asap}
\end{figure}

The training set consists of 7,336 structures, each having either 135 or 144 atoms. 
Amongst these, 3,483 liquid structures were generated by
running molecular dynamics simulations using both the MA and reMA potentials and selecting structures from the trajectory separated by a time interval that larger than the decorrelation time, and 3,853 solid structures were constructed by adding random displacements of atoms to the structures of known phases at different pressures.  The set includes a wide variety of geometries from cubic to triclinic at densities from 0.15~f.u./\AA$^3$ to 0.6~f.u./\AA$^3$, which is adequate to cover the range of pressures and temperatures we investigate.  A projection of the training set using principal component analysis is show in Fig~\ref{fig:asap}.

In the MLP, the atomic environments are described by a set of symmetry functions (SFs).
Initially a large, comprehensive set of SFs are generated with a maximum cutoff radius of $16r_\text{Bohr}$.
The SFs that make the most important contributions to modelling the interactions (without unnecessarily increasing computational cost) are selected using a decomposition algorithm similar to the CUR decomposition~\cite{Cersonsky2021,Imbalzano2018,Albert2015}.  
In addition to the CUR, the method is also biased to select an SF if it is strongly correlated with the magnitude of the forces in the dataset.  
We attempted training with 90, 117, and 150 unique SFs and found that using 90 SFs resulted in the best precision, due to overfitting when using greater numbers of SFs.

After training for 30 epochs using the Kalman Filter to update the 2,282 weights, we obtain the test and training errors which are summarised and compared with characteristic values for the training set in Table~\ref{tab:mlpErrors}.  The 30th epoch was chosen for subsequent calculations as it had the minimum RMS force error. 
\begin{table}[]
    \centering
    \begin{tabular}{c|c|c}
     					& Energy / eV~at$^{-1}$	& Forces / eV~\AA$^{-2}$ \\
    \hline %
    RMS Test Error			& 0.0110			& 0.725	 \\
    RMS Training Error		& 0.0099			& 0.713	 \\
    PBEsol Set SD 	& 0.38	 		& 3.14	 \\
    PBEsol-reMA Set SD 	& 0.23	 		& 1.21	 \\
    \end{tabular}
    \caption{
     Test and training RMSEs are compared with the standard deviations (SD) of the PBEsol training set before and after subtracting the contribution from the reMA potential (PBEsol-reMA).}
    \label{tab:mlpErrors}
\end{table}

It should be stressed that the purpose of the MLP we have trained is to provide a small correction to the behavior of the empirical reMA potential, and their effects must be superposed to produce accurate behavior.
Due to the reMA baseline, even for the regions not covered by the training set, the combined empirical potential and MLP will behave reasonably and remain stable.

\subsection{Enthalpies of TiO$_2$ predicted by the $\Delta$-learning potential}

We benchmark the accuracy of the $\Delta$-learning potential by comparing the enthalpy as a function of pressure at 0~K with that expected from DFT and that predicted by the MA potential (see Fig.~\ref{fig:H}).  
Although the relative stabilities of several metastable phases noticeably differ from the DFT reference, 
the $\Delta$-learning potential is able to correctly predict the relative enthalpies of the stable phases.
For instance, anatase is the stable phase at pressures up to 1~GPa, where baddeleyite becomes the stable phase until the compression reaches approximately 18~GPa, whereupon both it and OI are competitive.   Above 26~GPa cotunnite remains the stable phase. 
The enthalpy curves predicted by the $\Delta$-learning potential are considerably more similar to the DFT predictions than the predictions of the MA potential, which predicts pyrite to be the stable phase over most of the pressure range, when in fact it is approximately 500~meV/f.u. higher in enthalpy than the stable phase. 
Therefore, in what follows we will use the $\Delta$-learning potential to study the phase behaviors of the stable phases, as well as the transition processes between them.

\subsection{MD simulations using the $\Delta$-learning potential}

MD simulations using the $\Delta$-learning potential were carried out using LAMMPS~\cite{Plimpton1995} patched with N2P2~\cite{Singraber2019}.
The hybrid/overlay potential style in LAMMPS was used to combine the reMA component and the MLP correction.  The simulations were performed using a 1~fs timestep, with Nose-Hoover barostatting and canonical sampling thermostatting.  The simulation cell is allowed to relax such that all of the cell lengths and angles can vary independently.  Full details of all MD simulations and geometry-optimisations are given in the Appendix.

\section{Phase diagram}

The metastability of the TiO$_2$ phases at non-zero temperatures is determined by their relative chemical potentials, instead of the enthalpies at 0~K.  
We thus computed the Gibbs free energies via the thermodynamic integration method - the method is briefly outlined in the Appendix and discussed in detail in Ref.~\citenum{Cheng2018}.  The free energy is computed at intervals of 100~K and 10~GPa, then interpolated to determine the phase boundaries illustrated in Fig. 3.

We find that at low temperature and pressure anatase is the stable phase, in agreement with experimental and theoretical observations.  Experimentally, anatase is predicted to transform to rutile in the region of 900~K~\cite{Jamieson1969,Hanaor2011}, however we do not observe rutile as a stable polymorph due to the known issue of DFT over-estimating its enthalpy~\cite{Reinhardt2020}.  

The $\Delta$-learning potential predicts that anatase instead transforms to the higher entropy phase brookite at 500-600~K.  Che~\cite{Che2016} also predicts brookite is a stable polymorph, although several sources claim columbite to be the more stable~\cite{Reinhardt2019,Muscat2002}. 

Baddeleyite is predicted to be the stable phase over a wide region, however the OI phase is also very competitive in the high pressure part of this region, as would be expected from experiment.  At pressures above 45~GPa we observe cotunnite, as expected from both experiments and DFT~\cite{Reinhardt2019, Dubrovinsky2001}.  

In comparison, the MA potential does not predict baddeleyite to be the stable phase at any pressure and temperature.  Although the MA potential does correctly predict rutile as the stable phase at low pressure, anatase is not predicted to be stable, and both OI and columbite are only stable in much smaller $P$-$T$ regions than are found in experiment~\cite{Reinhardt2019}. 

\begin{figure}
    \centering
    \includegraphics[width=\linewidth]{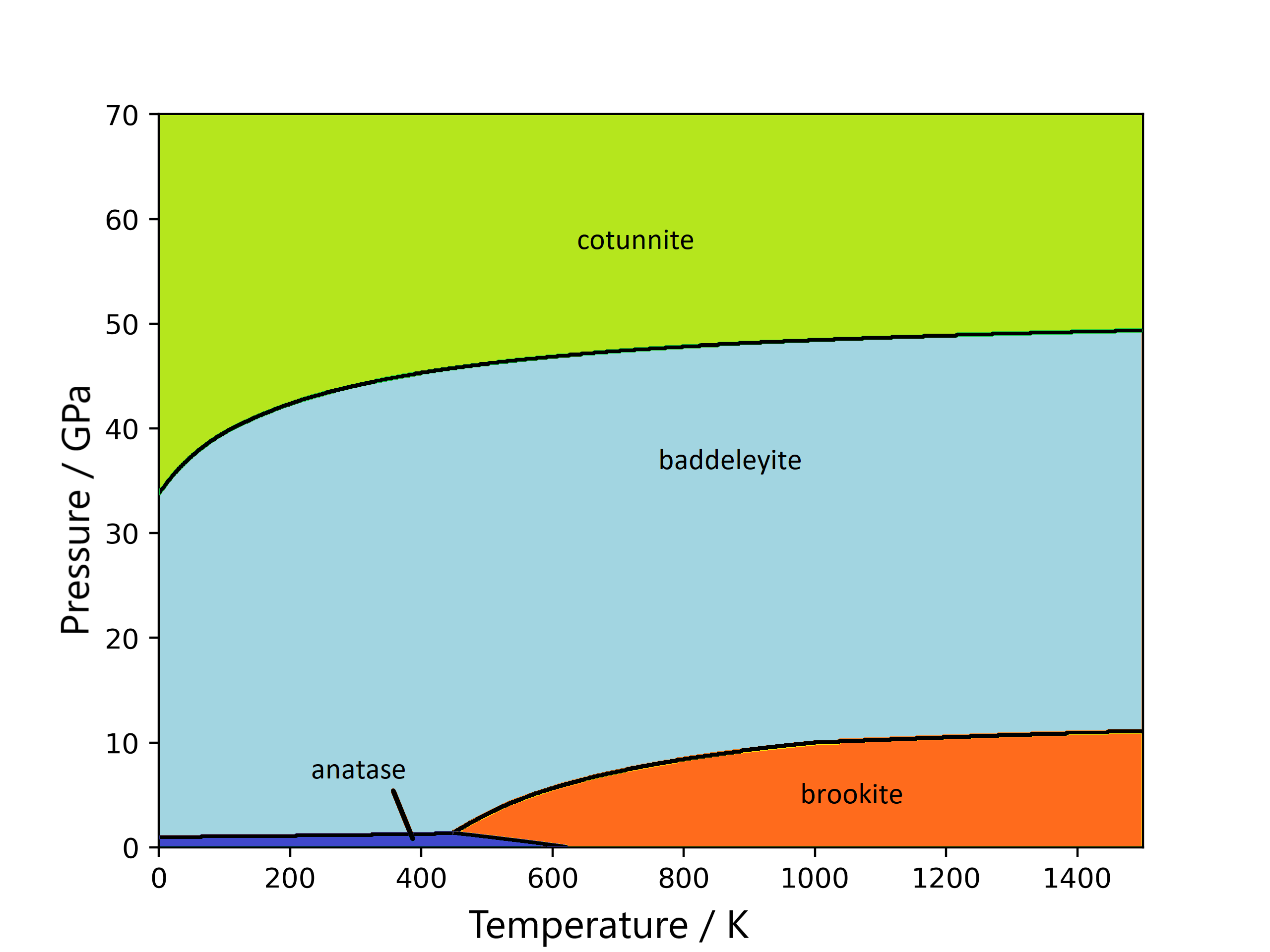}
    \caption{Phase diagram.  The phase with the minimum Gibbs free at each point in the parameter space is shown, and the boundaries are determined by interpolation.  Note that OI is also competitive with baddeleyite at higher pressures.}
    \label{fig:GvsT}
\end{figure}

\section{Solid-Solid Phase Transitions~\label{sec:ss}} %

In practise, high pressure phases such as cotunnite are usually made by compression of anatase or rutile using a diamond anvil and heating the sample either electrically or by laser, then decompressing back to ambient pressure and quenching~\cite{Swamy2002}.  However, different experiments have predicted different phases will be formed.  

Furthermore, it has not previously been possible to elucidate the transition mechanisms either experimentally - due to the difficulty of making in-situ measurements of very short-timescale processes - or theoretically, due to the computational cost of \textit{ab intio} methods and the poor accuracy of empirical potentials.  The $\Delta$-learning potential that we have developed is capable of correctly ordering the metastability of some of the phases, while being computationally cheap enough to be able to perform long timescale MD simulations.  

We demonstrate this by compressing anatase from 0 to 70~GPa over a period of 1~ns.  A timestep of 1~fs is used and the thermostat and barostat routines are the same as above.  Both the lengths and angles of the simulation cell are allowed to vary independently.  The simulation is repeated using three different system sizes, 128, 432, or 1024 TiO$_2$ formula units (f.u.), as well as 12 different random seeds for each size.  
We performed independent simulations at 300~K, 500~K, 800~K, 1000~K, 1500~K, and 2000~K.
The enthalpy and volume evolution of the systems during the compression at 300~K and 2000~K is given in Fig.~\ref{fig:anataseCompression},
and the rest of the results are in the Appendix.
For comparison, the results from the compression simulations of baddeleyite and cotunnite are also included.
In order to better distinguish between the different phases,
both enthalpy and volume are plotted as moving averages over 100 timesteps and the enthalpies are given relative to a linear fit of the enthalpy of cotunnite.
At 300~K using the small system size of 128 f.u. (the 12 sets of blue curves in Fig.~\ref{fig:anataseCompression}a), 
most (9 out of 12) simulations exhibited a two-step transition, with the first transition at about 20~GPa and the second one at about 40~GPa,
and amongst them 8 simulations reached the baddeleyite phase.
With the system size of 432 f.u. (the light blue curves in Fig.~\ref{fig:anataseCompression}a),
7 out of 12 show the same two-step transition, and 6 became baddeleyite.
With the 1024 f.u. simulation cell, 4 runs have the same transition, and all ended up as baddeleyite.
As such, the transition process is quite consistent at the three different system sizes.

To understand the two-step transition mechanism better,
we plot the displacement vectors of all the atoms in the small system along with the snapshots before and after the two transitions in Fig.~\ref{fig:displacements}a.
The displacement vectors show the atomic movement $\textbf{q}(t)-\textbf{q}(t_0)$, where $\textbf{q}(t_0)$ and $\textbf{q}(t)$ are the coordinates just before and after the transition. 
Fig.~\ref{fig:displacements}a shows that,
in both steps,
all the atoms follow a highly ordered and collective motion,
and the most prominent feature is the sliding of two of the \{010\} planes over each other.
Such transitions are analogous to the diffusionless Martensitic transition.

At 300~K and with all three different system sizes,
a one-step transition mechanism was sometimes observed,
during which the sharp change happens at about 20~GPa,
and the final structure has a density between that of baddeleyite and cotunnite,
and an enthalpy that is higher than baddeleyite.
We show the displacement vector analysis for this one-step transition in Fig.~\ref{fig:displacements}b.
It can be seen that the arrangement of Ti atoms in the ending configuration resembles baddeleyite,
but the oxygen atoms are less ordered.
The displacement vectors again show highly ordered and collective motion. 
Comparing the one-step transition with the first transformation in the aforementioned two-step process,
our results suggest that while the pressure at which this transition begins is consistent (about 20~GPa),
the product state of the phase transition is determined probabilistically between the baddeleyite and the oxygen-disordered baddeleyite-like state. 

Fig.~\ref{fig:anataseCompression}b shows the evolution of volume and enthalpy during the compression of anatase at a high temperature of 2000~K.
In this case,  a fraction of the end product from simulations with the small system size (the 12 sets of blue curves in Fig.~\ref{fig:anataseCompression}b) are the thermodynamically-stable cotunnite,
while the rest are in the oxygen-disordered baddeleyite-like state.
This is in contrast to the low temperature simulations at 300~K, where none of the products were cotunnite.
With larger system sizes, as shown using the light blue and the orange curves in Fig.~\ref{fig:anataseCompression}b,
all the final structures have the oxygen-disordered baddeleyite-like character.
This suggests that simulations run with smaller system sizes are more likely to end in the stable polymorph.

In all the simulation runs,
the volume and enthalpy curves show a sharp drop at about 20~GPa,
follows by gradual and somewhat stochastic changes in both volume and energy between about 20~GPa-30~GPa.
During the runs that ended up in cotunnite, there is also a third sharp transition at a pressure between 45~GPa and 55~GPa depending on the initial random seed of the simulation.  These transition pressures are slightly higher than predicted by the phase diagram, but this in not unexpected as the kinetic barriers must be overcome and the rate of compression is very high.
We show the displacement vector analysis for such three-step process in Fig.~\ref{fig:displacements}c.
For the first and the third transition,
the motion of both Ti and O atoms are sudden as well as highly-coordinated,
and the shuffling of the whole atomic planes are visible as well.
During the second transition, from $t=300$~ps to $t=400$~ps and pressure from about 20~GPa-30~GPa, however,
the displacements of the Ti atoms are relatively small,
while the O atoms undergo large displacements with seemingly random orientations.
These oxygen motions thus appear diffusive,
and such diffusion-like behaviors are typically associated with thermal activation.
If oxygen diffusion is indeed required in the anatase to cotunnite transformation process,
it would explain why the simulations run with smaller system sizes are more likely to end in this stable polymorph, 
and why this transformation is never seen in simulations at lower temperatures.

Experimental results for the compression of anatase are mixed and are dependent on the microstructure of the initial material.  For example, it has been shown that single crystal anatase transforms to columbite at 7~GPa~\cite{Arlt2000}, whereas in polycrystalline anatase this transformation is suppressed, and it instead transforms to baddeleyite when compressed to 18~GPa and heated to 850-900~K~\cite{Swamy2003}.  While we would expect our results to reflect the single crystal case, we find that the transformation to columbite is suppressed.  Nevertheless, the anatase-baddeleyite transition pressure is in good agreement and the temperature dependence of the transition is consistent with our suggestion that it may involve oxygen diffusion.

Under further compression, baddeleyite has been shown experimentally to transform to cotunnite when compressed to 61~GPa and heated to 1100~K~\cite{Dubrovinsky2001}.  Our  
simulations with the MLP show this transformation occurring only at high temperatures between 46 and 57~GPa.

\begin{figure*}
    \centering
    \includegraphics[width=0.8\linewidth]{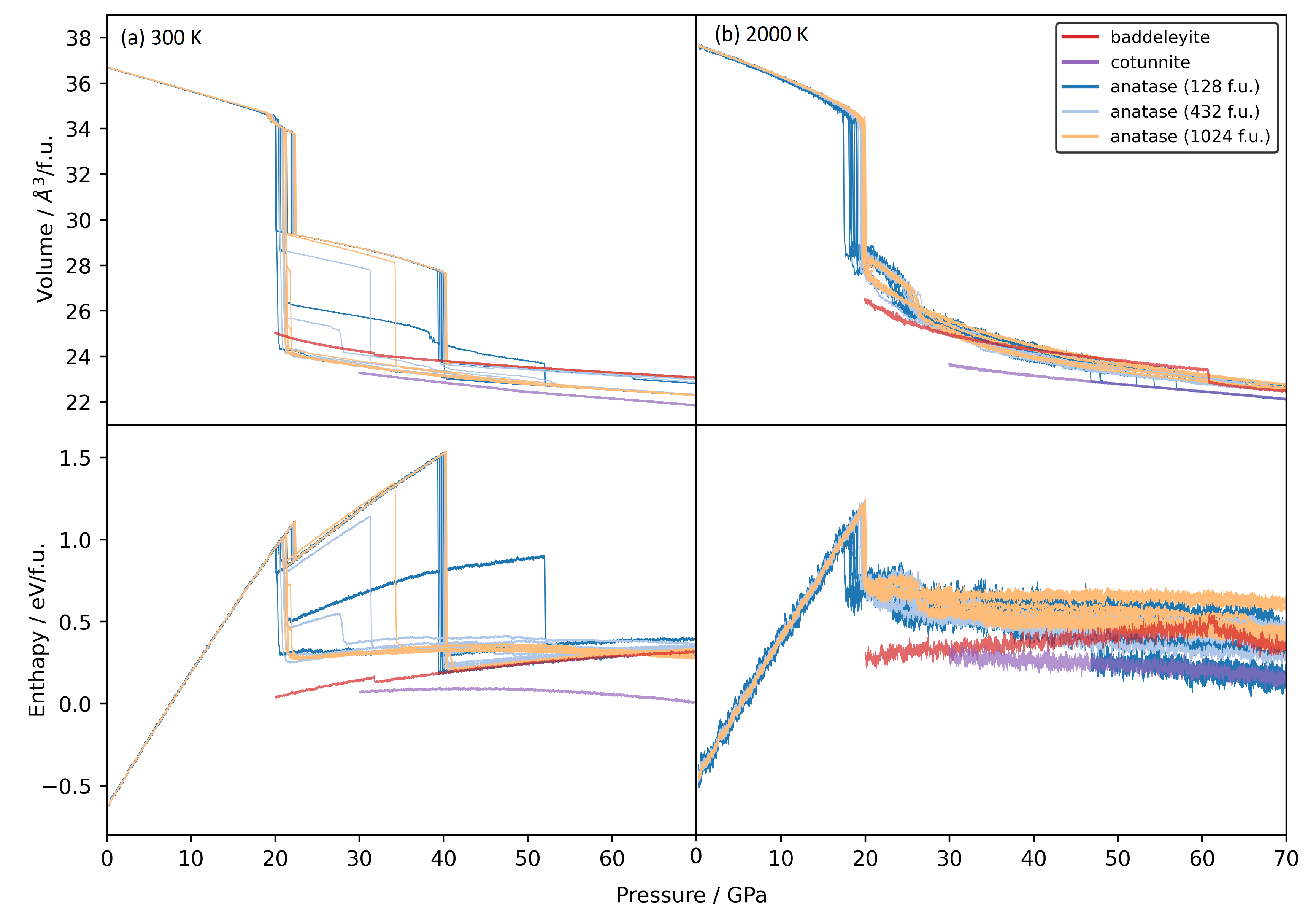}
    \caption{Volume and enthalpy over the course of rapid compression of anatase from 0 to 70~GPa over 1~ns at (a) 300~K and (b) 2000~K.  The compression was repeated 12 different times using different random seeds, for 3 different system sizes.  For reference, baddeleyite and cotunnite are also included.}
    \label{fig:anataseCompression}
\end{figure*}

\begin{figure*}
    \centering
    \includegraphics[width=0.9\linewidth]{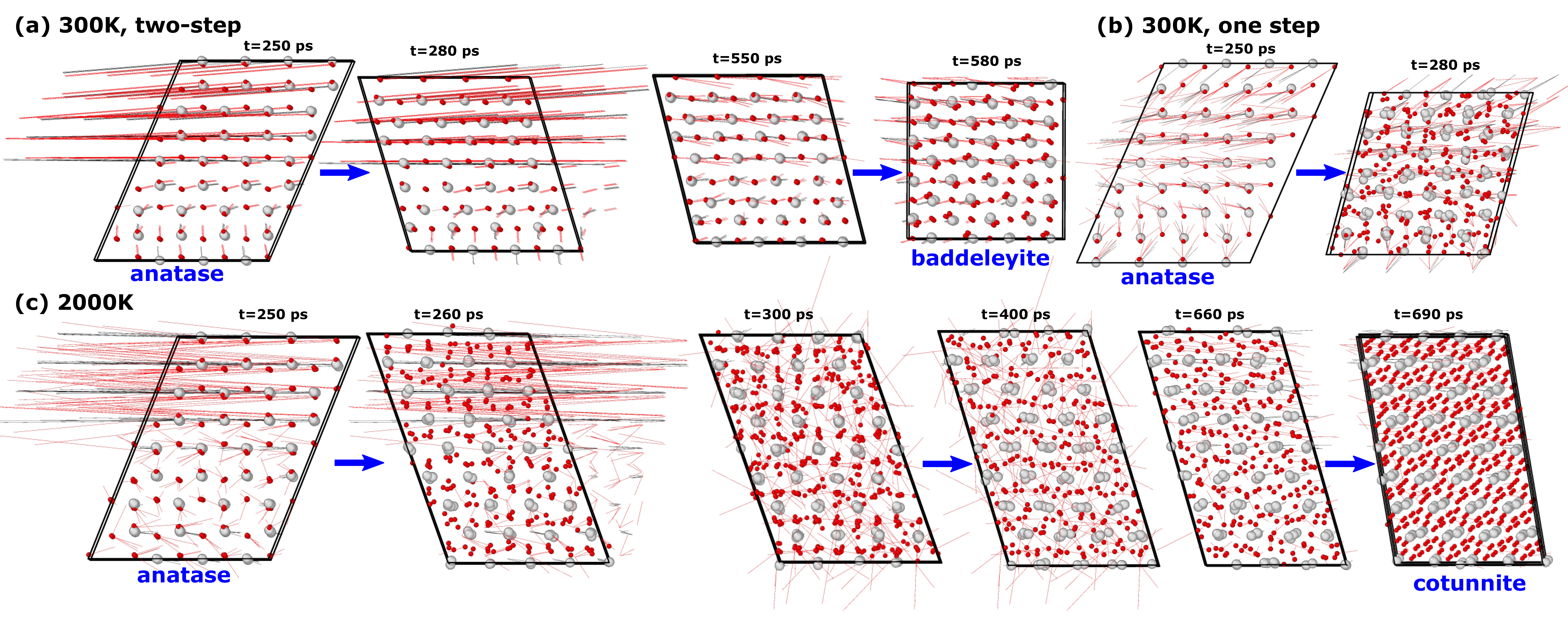}
    \caption{The displacement vector plots for the TiO$_2$ system before and after each transition during the rapid compression simulations (0~GPa at $t=0$~ps and 70~GPa and $t=1$~ns).
    The titanium atoms are plotted as gray spheres, and the associated displacement vectors are plotted as thin gray lines.
    The oxygen atoms and their displacements are both plotted in the red color.
    The snapshots of atomic coordinates were averaged over a time window of 2~ps.
    For each transition, both the starting and the ending configurations are provided, and the displacement vectors shown on top of both configurations are essentially the vectors that show the change of the positions of atoms.
    (a) shows a two-step transition between anatase and baddeleyite at 300~K.
    (b) illustrates an one-step transition between anatase and a baddeleyite-like phase with disordered oxygen atoms at 300~K.
    (c) depicts a three-step process for the transformation between anatase and cotunnite,
    with diffusive movement of oxygen atoms during the second step.
    }
    \label{fig:displacements}
\end{figure*}

\section{Conclusions} %

We have trained a stable, long-range MLP with significantly higher accuracy than empirical potentials for titanium dioxide using the $\Delta$-Learning method, where the training data consists of the difference between the energies and forces computed using DFT (using the PBEsol functional) and a reparametrised version of the MA potential (reMA).  The computational cost is $O(N)$ and approximately one order of magnitude higher than reMA, compared to $O(N^3)$ for DFT.

The relative enthalpy of the metastable phases shows good agreement with DFT at 0~K from 0-70~GPa, and predicts that the anatase, brookite, baddeleyite, and cotunnite phases would be the stable phase in different regions of $P$-$T$ space.

We have investigated the phase transformations that occur on compressing anatase from 0 to 70~GPa at selected temperatures ranging from 300~K to 2000~K.
We observed several distinct solid-solid transition mechanisms:
At low $T$, anatase can either undergo a two-step martensitic transformation to baddeleyite, with the first step involving the crystallographic planes gliding over each other, or a single-step transformation to an oxygen-disordered baddeleyite-like state.
At high $T$, anatase can transform into cotunnite via a thermally activated process that involves both gliding of the planes and 
diffusive movements of oxygen atoms.
The transition pressures for the anatase-baddeleyite and the baddeleyite-cotunnite transformations in our simulations agree well with experiments.

From the methodology side, future work could involve designing more sophisticated empirical potentials as the baseline of the $\Delta$-learning potential, which may include charge redistributions. 
Another direction is to use a higher-level electronic structure method as the reference for constructing MLPs, in order to avoid the known issues of DFT for the system of TiO$_2$.
For applications,
further work could be done using the MLP we have developed to investigate the many other phase transitions in titanium dioxide in order to develop more sustainable synthetic routes, in addition to further research into other complex phenomena such as polyamorphous phases~\cite{Deringer2021}, and surface physics.

\section{Appendix}

\subsection{Details on geometry optimisation} %

Starting with DFT-optimised structures of the known phases from Ref.~\citenum{Reinhardt2019} at 0~K and 0~GPa, the geometry is re-optimised using the MLP at 0~K and 0~GPa.  The box is allowed to relax by up to 10\% in volume to reach the target pressure and both the lengths and angles of the simulation cell are free to change during the relaxation.  Following the box relaxation, the quadratic conjugate gradient minimisation algorithm~\cite{Teter1989} is used to optimise the atomic positions.  
To avoid trapping in a shallow local minimum, the optimization is repeated four times, using sequentially smaller tolerances.  
The potential energy of the final structure is minimised to a precision of $10^{-10}$.  

The 0~K, 0~GPa is then used as the starting structure for geometry optimisation at 0~K, 5~GPa, and the structure is optimised in the same way.  This is repeated in 5~GPa intervals up to 70~GPa.  This is done for all phases in order to calculate the enthalpy curves shown in Fig.~\ref{fig:H}.

\subsection{Details on free energy calculations}

For each polymorph of TiO$_2$,
we first perform a molecular dynamics simulation at each pressure of interest at a low temperature $T_0=$100~K, starting with the corresponding structure optimised at 0~K.  
This is run for 30,000 $\times$ 1~fs timesteps in the isobaric-isothermal (NPT) ensemble using the Nose-Hoover barostat with a damping time of 500~fs.  
The required temperature is maintained using stochastic velocity rescaling thermostat~\cite{Bussi2007} with a damping time of 100~fs.  
The dimensions of the simulation cell are allowed to change anisotropically but the cell angles are fixed.  

The resulting geometries are optimised in a similar manner to the 0~K structures and their internal energies $U(T_0)$ are found.  Where a local minimum is found, the Hessian matrices of the structures are computed using the program \verb|i-PI|~\cite{Kapil2019,Ceriotti2014} using the finite-difference method.  The Hessian matrix enables us to construct a reference harmonic crystal, where the forces between the atoms are defined by the phonon modes.  
An MD simulation is run on the harmonic crystal in the isochoric-isothermal (NVT) ensemble to determine its internal energy $U_{har}(T_0)$.

Given the internal energies of the real and harmonic crystals, as well as the Helmholtz energy of the harmonic crystal $F_{har}(T_0)$ (which is the sum of the phonon energies) we can transform to the isobaric-isothermal (NPT) ensemble to determine the Gibbs energy using the equation,
\begin{equation}
    G(P,T_0) = F_{har}(V,T_0) - k_B T_0 \ln \left\langle \exp{\left( - \frac{U - U_{har}}{k_B T_0} \right)}\right\rangle_{V,T_0}.
\end{equation}

Further MD simulations at higher temperatures $200 \leq T_1 \leq 1500$~K are performed to determine the enthalpy of each polymorph under these conditions.  Thermodynamic integration is then used to determine the Gibbs energy at higher temperatures,
\begin{equation}
\frac{G(P,T_1)}{k_B T_1} = \frac{G(P,T_0)}{k_B T_0} - \int_{T_0}^{T_1} \frac{\langle H \rangle_{P,T}}{k_B T^2}dT.
\end{equation}

The measured properties, such as volume, enthalpy, etc., were extracted from the final 20\% of each MD simulation, when it has reached a stable equilibrium.  %

\begin{figure}[p]
    \centering
    \includegraphics[width=0.49\textwidth]{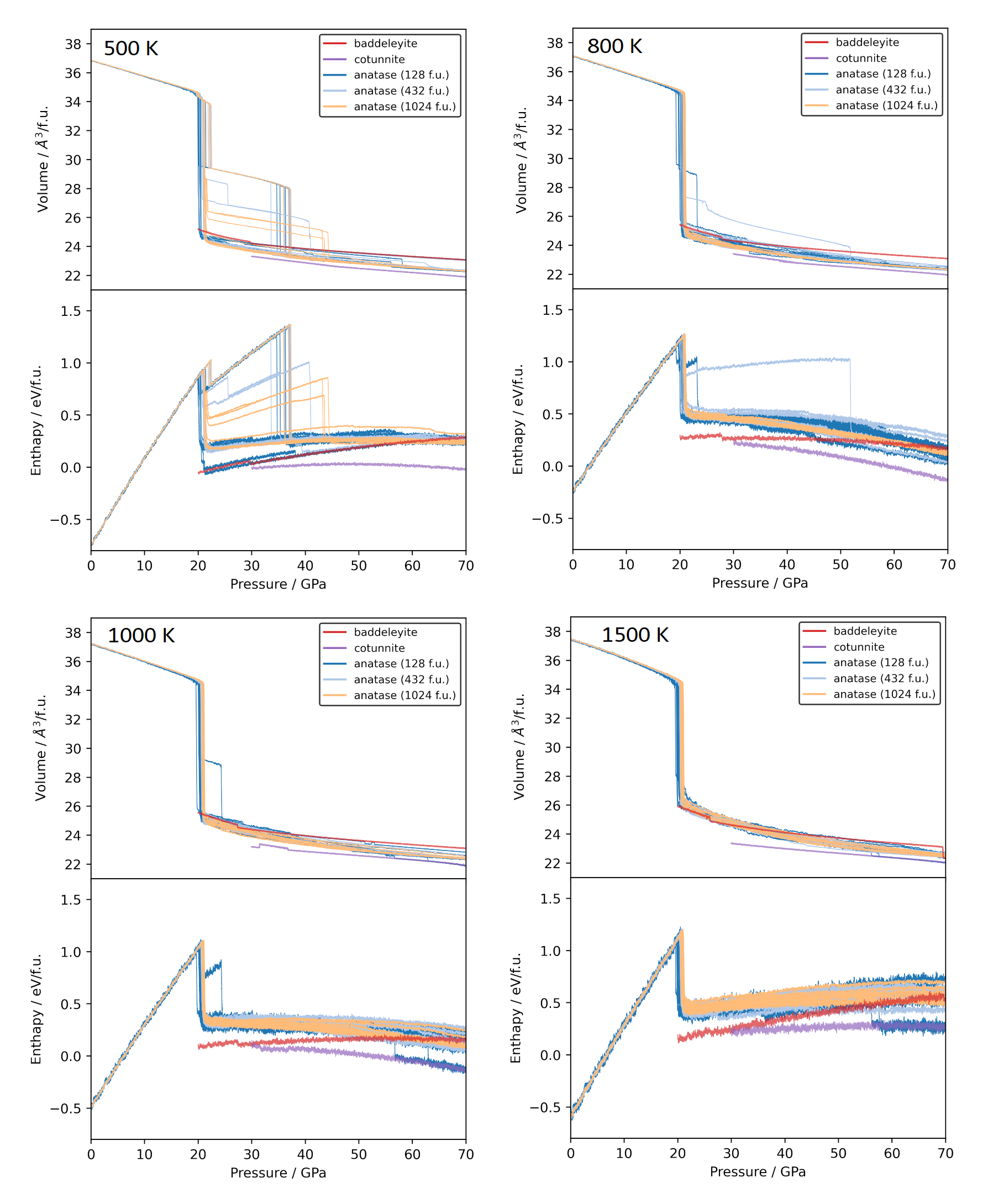}
    \caption{Enthalpy and volume curves for anatase as it is compressed from 0 - 70~GPa at $T =$~500, 800, 1000, 1500~K.  Cotunnite and baddeleyite are also shown for reference.}
    \label{fig:allCompressions}
\end{figure}

\subsection{More results on the anatase compression simulations}

In the main text, we analyze the simulations of  compressing anatase from 0 to 70~GPa over a period of 1~ns at 300~K and 2000~K (see Fig.~\ref{fig:anataseCompression}).
In Fig.~\ref{fig:allCompressions}, we show the analogous 
enthalpy and volume curves from simulations at $T =$ 500, 800, 1000, and 1500~K.
The observations in Sec.~\ref{sec:ss} are valid for these simulations at the intermediate temperatures:
the product state of the phase transition is determined probabilistically, and the final state is more likely to be cotunnite when the temperature is high.

\textbf{Acknowledgements}
JGL and BC acknowledges resources provided by the Cambridge Tier-2 system operated by the University of Cambridge Research Computing Service funded by EPSRC Tier-2 capital grant EP/P020259/1. 

\textbf{Data availability statement}
All original data generated for the study,
and the $\Delta$-learning potential for TiO$_2$ constructed in this study are in the %
repository \url{https://github.com/jacobglee1/tio2-mlp}


\begin{thebibliography}{64}%
\makeatletter
\providecommand \@ifxundefined [1]{%
 \@ifx{#1\undefined}
}%
\providecommand \@ifnum [1]{%
 \ifnum #1\expandafter \@firstoftwo
 \else \expandafter \@secondoftwo
 \fi
}%
\providecommand \@ifx [1]{%
 \ifx #1\expandafter \@firstoftwo
 \else \expandafter \@secondoftwo
 \fi
}%
\providecommand \natexlab [1]{#1}%
\providecommand \enquote  [1]{``#1''}%
\providecommand \bibnamefont  [1]{#1}%
\providecommand \bibfnamefont [1]{#1}%
\providecommand \citenamefont [1]{#1}%
\providecommand \href@noop [0]{\@secondoftwo}%
\providecommand \href [0]{\begingroup \@sanitize@url \@href}%
\providecommand \@href[1]{\@@startlink{#1}\@@href}%
\providecommand \@@href[1]{\endgroup#1\@@endlink}%
\providecommand \@sanitize@url [0]{\catcode `\\12\catcode `\$12\catcode
  `\&12\catcode `\#12\catcode `\^12\catcode `\_12\catcode `\%12\relax}%
\providecommand \@@startlink[1]{}%
\providecommand \@@endlink[0]{}%
\providecommand \url  [0]{\begingroup\@sanitize@url \@url }%
\providecommand \@url [1]{\endgroup\@href {#1}{\urlprefix }}%
\providecommand \urlprefix  [0]{URL }%
\providecommand \Eprint [0]{\href }%
\providecommand \doibase [0]{http://dx.doi.org/}%
\providecommand \selectlanguage [0]{\@gobble}%
\providecommand \bibinfo  [0]{\@secondoftwo}%
\providecommand \bibfield  [0]{\@secondoftwo}%
\providecommand \translation [1]{[#1]}%
\providecommand \BibitemOpen [0]{}%
\providecommand \bibitemStop [0]{}%
\providecommand \bibitemNoStop [0]{.\EOS\space}%
\providecommand \EOS [0]{\spacefactor3000\relax}%
\providecommand \BibitemShut  [1]{\csname bibitem#1\endcsname}%
\let\auto@bib@innerbib\@empty
%</preamble>
\bibitem [{\citenamefont {Yin}\ \emph {et~al.}(2013)\citenamefont {Yin},
  \citenamefont {Wu}, \citenamefont {Yang},\ and\ \citenamefont
  {Su}}]{Yin2013}%
  \BibitemOpen
  \bibfield  {author} {\bibinfo {author} {\bibfnamefont {Z.~F.}\ \bibnamefont
  {Yin}}, \bibinfo {author} {\bibfnamefont {L.}~\bibnamefont {Wu}}, \bibinfo
  {author} {\bibfnamefont {H.~G.}\ \bibnamefont {Yang}}, \ and\ \bibinfo
  {author} {\bibfnamefont {Y.~H.}\ \bibnamefont {Su}},\ }\bibfield  {title}
  {\enquote {\bibinfo {title} {Recent progress in biomedical applications of
  titanium dioxide},}\ }\href@noop {} {\bibfield  {journal} {\bibinfo
  {journal} {Physical chemistry chemical physics}\ }\textbf {\bibinfo {volume}
  {15}},\ \bibinfo {pages} {4844--4858} (\bibinfo {year} {2013})}\BibitemShut
  {NoStop}%
\bibitem [{\citenamefont {Mor}\ \emph {et~al.}(2006)\citenamefont {Mor},
  \citenamefont {Shankar}, \citenamefont {Paulose}, \citenamefont {Varghese},\
  and\ \citenamefont {Grimes}}]{Mor2006}%
  \BibitemOpen
  \bibfield  {author} {\bibinfo {author} {\bibfnamefont {G.~K.}\ \bibnamefont
  {Mor}}, \bibinfo {author} {\bibfnamefont {K.}~\bibnamefont {Shankar}},
  \bibinfo {author} {\bibfnamefont {M.}~\bibnamefont {Paulose}}, \bibinfo
  {author} {\bibfnamefont {O.~K.}\ \bibnamefont {Varghese}}, \ and\ \bibinfo
  {author} {\bibfnamefont {C.~A.}\ \bibnamefont {Grimes}},\ }\bibfield  {title}
  {\enquote {\bibinfo {title} {Use of highly-ordered tio2 nanotube arrays in
  dye-sensitized solar cells},}\ }\href {\doibase 10.1021/nl052099j} {\bibfield
   {journal} {\bibinfo  {journal} {Nano Letters}\ }\textbf {\bibinfo {volume}
  {6}},\ \bibinfo {pages} {215--218} (\bibinfo {year} {2006})},\ \bibinfo
  {note} {pMID: 16464037},\ \Eprint
  {http://arxiv.org/abs/https://doi.org/10.1021/nl052099j}
  {https://doi.org/10.1021/nl052099j} \BibitemShut {NoStop}%
\bibitem [{\citenamefont {Hagfeldt}\ and\ \citenamefont
  {Graetzel}(1995)}]{Hagfeldt1995}%
  \BibitemOpen
  \bibfield  {author} {\bibinfo {author} {\bibfnamefont {A.}~\bibnamefont
  {Hagfeldt}}\ and\ \bibinfo {author} {\bibfnamefont {M.}~\bibnamefont
  {Graetzel}},\ }\bibfield  {title} {\enquote {\bibinfo {title} {Light-induced
  redox reactions in nanocrystalline systems},}\ }\href {\doibase
  10.1021/cr00033a003} {\bibfield  {journal} {\bibinfo  {journal} {Chemical
  Reviews}\ }\textbf {\bibinfo {volume} {95}},\ \bibinfo {pages} {49--68}
  (\bibinfo {year} {1995})},\ \Eprint
  {http://arxiv.org/abs/https://doi.org/10.1021/cr00033a003}
  {https://doi.org/10.1021/cr00033a003} \BibitemShut {NoStop}%
\bibitem [{\citenamefont {Fujishima}\ and\ \citenamefont
  {Honda}(1972)}]{Fujishima1972}%
  \BibitemOpen
  \bibfield  {author} {\bibinfo {author} {\bibfnamefont {A.}~\bibnamefont
  {Fujishima}}\ and\ \bibinfo {author} {\bibfnamefont {K.}~\bibnamefont
  {Honda}},\ }\bibfield  {title} {\enquote {\bibinfo {title} {Electrochemical
  photolysis of water at a semiconductor electrode},}\ }\href@noop {}
  {\bibfield  {journal} {\bibinfo  {journal} {nature}\ }\textbf {\bibinfo
  {volume} {238}},\ \bibinfo {pages} {37--38} (\bibinfo {year}
  {1972})}\BibitemShut {NoStop}%
\bibitem [{\citenamefont {Jeong}, \citenamefont {Youn},\ and\ \citenamefont
  {Jeon}(2020)}]{Jeong2020}%
  \BibitemOpen
  \bibfield  {author} {\bibinfo {author} {\bibfnamefont {S.}~\bibnamefont
  {Jeong}}, \bibinfo {author} {\bibfnamefont {J.-S.}\ \bibnamefont {Youn}}, \
  and\ \bibinfo {author} {\bibfnamefont {K.-J.}\ \bibnamefont {Jeon}},\
  }\bibfield  {title} {\enquote {\bibinfo {title} {Titanium dioxide-coated
  copper electrodes for hydrogen production by water splitting},}\ }\href
  {\doibase https://doi.org/10.1016/j.ijhydene.2019.07.239} {\bibfield
  {journal} {\bibinfo  {journal} {International Journal of Hydrogen Energy}\
  }\textbf {\bibinfo {volume} {45}},\ \bibinfo {pages} {24037 -- 24044}
  (\bibinfo {year} {2020})},\ \bibinfo {note} {proceedings of 5th International
  Conference on Nanotechnology, Nanomaterials \& Thin Films for Energy
  Applications – 2018}\BibitemShut {NoStop}%
\bibitem [{\citenamefont {Dubrovinsky}\ \emph {et~al.}(2001)\citenamefont
  {Dubrovinsky}, \citenamefont {Dubrovinskaia}, \citenamefont {Swamy},
  \citenamefont {Muscat}, \citenamefont {Harrison}, \citenamefont {Ahuja},
  \citenamefont {Holm},\ and\ \citenamefont {Johansson}}]{Dubrovinsky2001}%
  \BibitemOpen
  \bibfield  {author} {\bibinfo {author} {\bibfnamefont {L.~S.}\ \bibnamefont
  {Dubrovinsky}}, \bibinfo {author} {\bibfnamefont {N.~A.}\ \bibnamefont
  {Dubrovinskaia}}, \bibinfo {author} {\bibfnamefont {V.}~\bibnamefont
  {Swamy}}, \bibinfo {author} {\bibfnamefont {J.}~\bibnamefont {Muscat}},
  \bibinfo {author} {\bibfnamefont {N.~M.}\ \bibnamefont {Harrison}}, \bibinfo
  {author} {\bibfnamefont {R.}~\bibnamefont {Ahuja}}, \bibinfo {author}
  {\bibfnamefont {B.}~\bibnamefont {Holm}}, \ and\ \bibinfo {author}
  {\bibfnamefont {B.}~\bibnamefont {Johansson}},\ }\bibfield  {title} {\enquote
  {\bibinfo {title} {The hardest known oxide},}\ }\href {\doibase
  10.1038/35070650} {\bibfield  {journal} {\bibinfo  {journal} {Nature}\
  }\textbf {\bibinfo {volume} {410}},\ \bibinfo {pages} {653--654} (\bibinfo
  {year} {2001})}\BibitemShut {NoStop}%
\bibitem [{\citenamefont {Al-Khatatbeh}, \citenamefont {Lee},\ and\
  \citenamefont {Kiefer}(2009)}]{AlKhatatbeh2009}%
  \BibitemOpen
  \bibfield  {author} {\bibinfo {author} {\bibfnamefont {Y.}~\bibnamefont
  {Al-Khatatbeh}}, \bibinfo {author} {\bibfnamefont {K.~K.~M.}\ \bibnamefont
  {Lee}}, \ and\ \bibinfo {author} {\bibfnamefont {B.}~\bibnamefont {Kiefer}},\
  }\bibfield  {title} {\enquote {\bibinfo {title} {High-pressure behavior of
  ${\text{tio}}_{2}$ as determined by experiment and theory},}\ }\href
  {\doibase 10.1103/PhysRevB.79.134114} {\bibfield  {journal} {\bibinfo
  {journal} {Phys. Rev. B}\ }\textbf {\bibinfo {volume} {79}},\ \bibinfo
  {pages} {134114} (\bibinfo {year} {2009})}\BibitemShut {NoStop}%
\bibitem [{\citenamefont {Olsen}, \citenamefont {Gerward},\ and\ \citenamefont
  {Jiang}(1999)}]{Olsen1999}%
  \BibitemOpen
  \bibfield  {author} {\bibinfo {author} {\bibfnamefont {J.}~\bibnamefont
  {Olsen}}, \bibinfo {author} {\bibfnamefont {L.}~\bibnamefont {Gerward}}, \
  and\ \bibinfo {author} {\bibfnamefont {J.}~\bibnamefont {Jiang}},\ }\bibfield
   {title} {\enquote {\bibinfo {title} {On the rutile/$\alpha$-pbo2-type phase
  boundary of tio2},}\ }\href {\doibase
  https://doi.org/10.1016/S0022-3697(98)00274-1} {\bibfield  {journal}
  {\bibinfo  {journal} {Journal of Physics and Chemistry of Solids}\ }\textbf
  {\bibinfo {volume} {60}},\ \bibinfo {pages} {229 -- 233} (\bibinfo {year}
  {1999})}\BibitemShut {NoStop}%
\bibitem [{\citenamefont {Dekura}\ \emph {et~al.}(2011)\citenamefont {Dekura},
  \citenamefont {Tsuchiya}, \citenamefont {Kuwayama},\ and\ \citenamefont
  {Tsuchiya}}]{Dekura2011}%
  \BibitemOpen
  \bibfield  {author} {\bibinfo {author} {\bibfnamefont {H.}~\bibnamefont
  {Dekura}}, \bibinfo {author} {\bibfnamefont {T.}~\bibnamefont {Tsuchiya}},
  \bibinfo {author} {\bibfnamefont {Y.}~\bibnamefont {Kuwayama}}, \ and\
  \bibinfo {author} {\bibfnamefont {J.}~\bibnamefont {Tsuchiya}},\ }\bibfield
  {title} {\enquote {\bibinfo {title} {Theoretical and experimental evidence
  for a new post-cotunnite phase of titanium dioxide with significant optical
  absorption},}\ }\href {\doibase 10.1103/PhysRevLett.107.045701} {\bibfield
  {journal} {\bibinfo  {journal} {Phys.\ Rev.\ Lett.}\ }\textbf {\bibinfo
  {volume} {107}},\ \bibinfo {pages} {045701} (\bibinfo {year}
  {2011})}\BibitemShut {NoStop}%
\bibitem [{\citenamefont {Nishio-Hamane}\ \emph {et~al.}(2010)\citenamefont
  {Nishio-Hamane}, \citenamefont {Shimizu}, \citenamefont {Nakahira},
  \citenamefont {Niwa}, \citenamefont {Sano-Furukawa}, \citenamefont {Okada},
  \citenamefont {Yagi},\ and\ \citenamefont {Kikegawa}}]{NishioHamane2010}%
  \BibitemOpen
  \bibfield  {author} {\bibinfo {author} {\bibfnamefont {D.}~\bibnamefont
  {Nishio-Hamane}}, \bibinfo {author} {\bibfnamefont {A.}~\bibnamefont
  {Shimizu}}, \bibinfo {author} {\bibfnamefont {R.}~\bibnamefont {Nakahira}},
  \bibinfo {author} {\bibfnamefont {K.}~\bibnamefont {Niwa}}, \bibinfo {author}
  {\bibfnamefont {A.}~\bibnamefont {Sano-Furukawa}}, \bibinfo {author}
  {\bibfnamefont {T.}~\bibnamefont {Okada}}, \bibinfo {author} {\bibfnamefont
  {T.}~\bibnamefont {Yagi}}, \ and\ \bibinfo {author} {\bibfnamefont
  {T.}~\bibnamefont {Kikegawa}},\ }\bibfield  {title} {\enquote {\bibinfo
  {title} {The stability and equation of state for the cotunnite phase of
  TiO2 up to {\SI{70}{\giga\pascal}}},}\ }\href {\doibase
  10.1007/s00269-009-0316-0} {\bibfield  {journal} {\bibinfo  {journal} {Phys.\
  Chem.\ Miner.}\ }\textbf {\bibinfo {volume} {37}},\ \bibinfo {pages}
  {129--136} (\bibinfo {year} {2010})}\BibitemShut {NoStop}%
\bibitem [{\citenamefont {Hanaor}\ and\ \citenamefont
  {Sorrell}(2011)}]{Hanaor2011}%
  \BibitemOpen
  \bibfield  {author} {\bibinfo {author} {\bibfnamefont {D.~A.}\ \bibnamefont
  {Hanaor}}\ and\ \bibinfo {author} {\bibfnamefont {C.~C.}\ \bibnamefont
  {Sorrell}},\ }\bibfield  {title} {\enquote {\bibinfo {title} {Review of the
  anatase to rutile phase transformation},}\ }\href@noop {} {\bibfield
  {journal} {\bibinfo  {journal} {Journal of Materials science}\ }\textbf
  {\bibinfo {volume} {46}},\ \bibinfo {pages} {855--874} (\bibinfo {year}
  {2011})}\BibitemShut {NoStop}%
\bibitem [{\citenamefont {Liu}\ \emph {et~al.}(2015)\citenamefont {Liu},
  \citenamefont {Ran}, \citenamefont {Liu},\ and\ \citenamefont
  {Liu}}]{Liu2015}%
  \BibitemOpen
  \bibfield  {author} {\bibinfo {author} {\bibfnamefont {Q.-J.}\ \bibnamefont
  {Liu}}, \bibinfo {author} {\bibfnamefont {Z.}~\bibnamefont {Ran}}, \bibinfo
  {author} {\bibfnamefont {F.-S.}\ \bibnamefont {Liu}}, \ and\ \bibinfo
  {author} {\bibfnamefont {Z.-T.}\ \bibnamefont {Liu}},\ }\bibfield  {title}
  {\enquote {\bibinfo {title} {Phase transitions and mechanical stability of
  {{TiO2}} polymorphs under high pressure},}\ }\href {\doibase
  10.1016/j.jallcom.2015.01.085} {\bibfield  {journal} {\bibinfo  {journal}
  {J.\ Alloys Compd.}\ }\textbf {\bibinfo {volume} {631}},\ \bibinfo {pages}
  {192--201} (\bibinfo {year} {2015})}\BibitemShut {NoStop}%
\bibitem [{\citenamefont {Gupta}\ and\ \citenamefont
  {Tripathi}(2011)}]{Gupta2011}%
  \BibitemOpen
  \bibfield  {author} {\bibinfo {author} {\bibfnamefont {S.~M.}\ \bibnamefont
  {Gupta}}\ and\ \bibinfo {author} {\bibfnamefont {M.}~\bibnamefont
  {Tripathi}},\ }\bibfield  {title} {\enquote {\bibinfo {title} {A review of
  tio 2 nanoparticles},}\ }\href@noop {} {\bibfield  {journal} {\bibinfo
  {journal} {chinese science bulletin}\ }\textbf {\bibinfo {volume} {56}},\
  \bibinfo {pages} {1639--1657} (\bibinfo {year} {2011})}\BibitemShut {NoStop}%
\bibitem [{\citenamefont {Dachille}, \citenamefont {Simons},\ and\
  \citenamefont {Roy}(1968)}]{Dachille1968}%
  \BibitemOpen
  \bibfield  {author} {\bibinfo {author} {\bibfnamefont {F.}~\bibnamefont
  {Dachille}}, \bibinfo {author} {\bibfnamefont {P.~Y.}\ \bibnamefont
  {Simons}}, \ and\ \bibinfo {author} {\bibfnamefont {R.}~\bibnamefont {Roy}},\
  }\bibfield  {title} {\enquote {\bibinfo {title} {{Pressure-temperature
  studies of anatase, brookite, rutile and TiO2-II}},}\ }\href@noop {}
  {\bibfield  {journal} {\bibinfo  {journal} {American Mineralogist}\ }\textbf
  {\bibinfo {volume} {53}},\ \bibinfo {pages} {1929--1939} (\bibinfo {year}
  {1968})},\ \Eprint
  {http://arxiv.org/abs/https://pubs.geoscienceworld.org/msa/ammin/article-pdf/53/11-12/1929/4256099/am-1968-1929.pdf}
  {https://pubs.geoscienceworld.org/msa/ammin/article-pdf/53/11-12/1929/4256099/am-1968-1929.pdf}
  \BibitemShut {NoStop}%
\bibitem [{\citenamefont {Léger}\ \emph {et~al.}(1993)\citenamefont {Léger},
  \citenamefont {Haines}, \citenamefont {Atouf},\ and\ \citenamefont
  {Tomaszewski}}]{Leger1993}%
  \BibitemOpen
  \bibfield  {author} {\bibinfo {author} {\bibfnamefont {J.}~\bibnamefont
  {Léger}}, \bibinfo {author} {\bibfnamefont {J.}~\bibnamefont {Haines}},
  \bibinfo {author} {\bibfnamefont {A.}~\bibnamefont {Atouf}}, \ and\ \bibinfo
  {author} {\bibfnamefont {P.}~\bibnamefont {Tomaszewski}},\ }\href@noop {}
  {\emph {\bibinfo {title} {High-Pressure Science and Technology}}}\ (\bibinfo
  {publisher} {American Institute of Physics, New York},\ \bibinfo {year}
  {1993})\BibitemShut {NoStop}%
\bibitem [{\citenamefont {Haines}\ and\ \citenamefont
  {Léger}(1993)}]{Haines1993}%
  \BibitemOpen
  \bibfield  {author} {\bibinfo {author} {\bibfnamefont {J.}~\bibnamefont
  {Haines}}\ and\ \bibinfo {author} {\bibfnamefont {J.}~\bibnamefont
  {Léger}},\ }\bibfield  {title} {\enquote {\bibinfo {title} {X-ray
  diffraction study of tio2 up to 49 gpa},}\ }\href {\doibase
  https://doi.org/10.1016/0921-4526(93)90025-2} {\bibfield  {journal} {\bibinfo
   {journal} {Physica B: Condensed Matter}\ }\textbf {\bibinfo {volume}
  {192}},\ \bibinfo {pages} {233--237} (\bibinfo {year} {1993})}\BibitemShut
  {NoStop}%
\bibitem [{\citenamefont {Sato}\ \emph {et~al.}(1991)\citenamefont {Sato},
  \citenamefont {Endo}, \citenamefont {Sugiyama}, \citenamefont {Kikegawa},
  \citenamefont {Shimomura},\ and\ \citenamefont {Kusaba}}]{Sato1991}%
  \BibitemOpen
  \bibfield  {author} {\bibinfo {author} {\bibfnamefont {H.}~\bibnamefont
  {Sato}}, \bibinfo {author} {\bibfnamefont {S.}~\bibnamefont {Endo}}, \bibinfo
  {author} {\bibfnamefont {M.}~\bibnamefont {Sugiyama}}, \bibinfo {author}
  {\bibfnamefont {T.}~\bibnamefont {Kikegawa}}, \bibinfo {author}
  {\bibfnamefont {O.}~\bibnamefont {Shimomura}}, \ and\ \bibinfo {author}
  {\bibfnamefont {K.}~\bibnamefont {Kusaba}},\ }\bibfield  {title} {\enquote
  {\bibinfo {title} {Baddeleyite-type high-pressure phase of TiO2},}\
  }\href {\doibase 10.1126/science.251.4995.786} {\bibfield  {journal}
  {\bibinfo  {journal} {Science}\ }\textbf {\bibinfo {volume} {251}},\ \bibinfo
  {pages} {786--788} (\bibinfo {year} {1991})}\BibitemShut {NoStop}%
\bibitem [{\citenamefont {Kalaiarasi}\ \emph {et~al.}(2018)\citenamefont
  {Kalaiarasi}, \citenamefont {Sivakumar}, \citenamefont {Dhas},\ and\
  \citenamefont {Jose}}]{Kalaiarasi2018}%
  \BibitemOpen
  \bibfield  {author} {\bibinfo {author} {\bibfnamefont {S.}~\bibnamefont
  {Kalaiarasi}}, \bibinfo {author} {\bibfnamefont {A.}~\bibnamefont
  {Sivakumar}}, \bibinfo {author} {\bibfnamefont {S.~M.~B.}\ \bibnamefont
  {Dhas}}, \ and\ \bibinfo {author} {\bibfnamefont {M.}~\bibnamefont {Jose}},\
  }\bibfield  {title} {\enquote {\bibinfo {title} {Shock wave induced anatase
  to rutile tio2 phase transition using pressure driven shock tube},}\
  }\href@noop {} {\bibfield  {journal} {\bibinfo  {journal} {Materials
  Letters}\ }\textbf {\bibinfo {volume} {219}},\ \bibinfo {pages} {72--75}
  (\bibinfo {year} {2018})}\BibitemShut {NoStop}%
\bibitem [{\citenamefont {Dauksta}\ \emph {et~al.}(2019)\citenamefont
  {Dauksta}, \citenamefont {Medvids}, \citenamefont {Onufrijevs}, \citenamefont
  {Shimomura}, \citenamefont {Fukuda},\ and\ \citenamefont
  {Murakami}}]{Dauksta2019}%
  \BibitemOpen
  \bibfield  {author} {\bibinfo {author} {\bibfnamefont {E.}~\bibnamefont
  {Dauksta}}, \bibinfo {author} {\bibfnamefont {A.}~\bibnamefont {Medvids}},
  \bibinfo {author} {\bibfnamefont {P.}~\bibnamefont {Onufrijevs}}, \bibinfo
  {author} {\bibfnamefont {M.}~\bibnamefont {Shimomura}}, \bibinfo {author}
  {\bibfnamefont {Y.}~\bibnamefont {Fukuda}}, \ and\ \bibinfo {author}
  {\bibfnamefont {K.}~\bibnamefont {Murakami}},\ }\bibfield  {title} {\enquote
  {\bibinfo {title} {Laser-induced crystalline phase transition from rutile to
  anatase of niobium doped tio2},}\ }\href@noop {} {\bibfield  {journal}
  {\bibinfo  {journal} {Current Applied Physics}\ }\textbf {\bibinfo {volume}
  {19}},\ \bibinfo {pages} {351--355} (\bibinfo {year} {2019})}\BibitemShut
  {NoStop}%
\bibitem [{\citenamefont {Muscat}, \citenamefont {Swamy},\ and\ \citenamefont
  {Harrison}(2002)}]{Muscat2002}%
  \BibitemOpen
  \bibfield  {author} {\bibinfo {author} {\bibfnamefont {J.}~\bibnamefont
  {Muscat}}, \bibinfo {author} {\bibfnamefont {V.}~\bibnamefont {Swamy}}, \
  and\ \bibinfo {author} {\bibfnamefont {N.~M.}\ \bibnamefont {Harrison}},\
  }\bibfield  {title} {\enquote {\bibinfo {title} {First-principles
  calculations of the phase stability of {{\ce{TiO2}}}},}\ }\href {\doibase
  10.1103/PhysRevB.65.224112} {\bibfield  {journal} {\bibinfo  {journal}
  {Phys.\ Rev.~B}\ }\textbf {\bibinfo {volume} {65}},\ \bibinfo {pages}
  {224112} (\bibinfo {year} {2002})}\BibitemShut {NoStop}%
\bibitem [{\citenamefont {Luo}\ \emph {et~al.}(2016)\citenamefont {Luo},
  \citenamefont {Benali}, \citenamefont {Shulenburger}, \citenamefont {Krogel},
  \citenamefont {Heinonen},\ and\ \citenamefont {Kent}}]{Luo2016}%
  \BibitemOpen
  \bibfield  {author} {\bibinfo {author} {\bibfnamefont {Y.}~\bibnamefont
  {Luo}}, \bibinfo {author} {\bibfnamefont {A.}~\bibnamefont {Benali}},
  \bibinfo {author} {\bibfnamefont {L.}~\bibnamefont {Shulenburger}}, \bibinfo
  {author} {\bibfnamefont {J.~T.}\ \bibnamefont {Krogel}}, \bibinfo {author}
  {\bibfnamefont {O.}~\bibnamefont {Heinonen}}, \ and\ \bibinfo {author}
  {\bibfnamefont {P.~R.~C.}\ \bibnamefont {Kent}},\ }\bibfield  {title}
  {\enquote {\bibinfo {title} {Phase stability of TiO2 polymorphs from
  diffusion {Q}uantum {M}onte {C}arlo},}\ }\href {\doibase
  10.1088/1367-2630/18/11/113049} {\bibfield  {journal} {\bibinfo  {journal}
  {New J.\ Phys.}\ }\textbf {\bibinfo {volume} {18}},\ \bibinfo {pages}
  {113049} (\bibinfo {year} {2016})}\BibitemShut {NoStop}%
\bibitem [{\citenamefont {Mei}\ \emph {et~al.}(2014)\citenamefont {Mei},
  \citenamefont {Wang}, \citenamefont {Shang},\ and\ \citenamefont
  {Liu}}]{Mei2014}%
  \BibitemOpen
  \bibfield  {author} {\bibinfo {author} {\bibfnamefont {Z.-G.}\ \bibnamefont
  {Mei}}, \bibinfo {author} {\bibfnamefont {Y.}~\bibnamefont {Wang}}, \bibinfo
  {author} {\bibfnamefont {S.}~\bibnamefont {Shang}}, \ and\ \bibinfo {author}
  {\bibfnamefont {Z.-K.}\ \bibnamefont {Liu}},\ }\bibfield  {title} {\enquote
  {\bibinfo {title} {First-principles study of the mechanical properties and
  phase stability of TiO2},}\ }\href {\doibase
  10.1016/j.commatsci.2013.11.020} {\bibfield  {journal} {\bibinfo  {journal}
  {Comput.\ Mater.\ Sci.}\ }\textbf {\bibinfo {volume} {83}},\ \bibinfo {pages}
  {114--119} (\bibinfo {year} {2014})}\BibitemShut {NoStop}%
\bibitem [{\citenamefont {Samat}\ \emph {et~al.}(2016)\citenamefont {Samat},
  \citenamefont {Ali}, \citenamefont {Taib}, \citenamefont {Hassan},\ and\
  \citenamefont {Yahya}}]{Samat2016}%
  \BibitemOpen
  \bibfield  {author} {\bibinfo {author} {\bibfnamefont {M.}~\bibnamefont
  {Samat}}, \bibinfo {author} {\bibfnamefont {A.}~\bibnamefont {Ali}}, \bibinfo
  {author} {\bibfnamefont {M.}~\bibnamefont {Taib}}, \bibinfo {author}
  {\bibfnamefont {O.}~\bibnamefont {Hassan}}, \ and\ \bibinfo {author}
  {\bibfnamefont {M.}~\bibnamefont {Yahya}},\ }\bibfield  {title} {\enquote
  {\bibinfo {title} {Hubbard u calculations on optical properties of 3d
  transition metal oxide tio2},}\ }\href@noop {} {\bibfield  {journal}
  {\bibinfo  {journal} {Results in physics}\ }\textbf {\bibinfo {volume} {6}},\
  \bibinfo {pages} {891--896} (\bibinfo {year} {2016})}\BibitemShut {NoStop}%
\bibitem [{\citenamefont {Dharmale}\ \emph {et~al.}(2020)\citenamefont
  {Dharmale}, \citenamefont {Chaudhury}, \citenamefont {Mahamune},\ and\
  \citenamefont {Dash}}]{Dharmale2020}%
  \BibitemOpen
  \bibfield  {author} {\bibinfo {author} {\bibfnamefont {N.}~\bibnamefont
  {Dharmale}}, \bibinfo {author} {\bibfnamefont {S.}~\bibnamefont {Chaudhury}},
  \bibinfo {author} {\bibfnamefont {R.}~\bibnamefont {Mahamune}}, \ and\
  \bibinfo {author} {\bibfnamefont {D.}~\bibnamefont {Dash}},\ }\bibfield
  {title} {\enquote {\bibinfo {title} {Comparative study on structural,
  electronic, optical and mechanical properties of normal and high pressure
  phases titanium dioxide using dft},}\ }\href@noop {} {\bibfield  {journal}
  {\bibinfo  {journal} {Materials Research Express}\ }\textbf {\bibinfo
  {volume} {7}},\ \bibinfo {pages} {054004} (\bibinfo {year}
  {2020})}\BibitemShut {NoStop}%
\bibitem [{\citenamefont {Arroyo-de Dompablo}, \citenamefont
  {Morales-Garc{\'\i}a},\ and\ \citenamefont {Taravillo}(2011)}]{Arroyo2011}%
  \BibitemOpen
  \bibfield  {author} {\bibinfo {author} {\bibfnamefont {M.}~\bibnamefont
  {Arroyo-de Dompablo}}, \bibinfo {author} {\bibfnamefont {A.}~\bibnamefont
  {Morales-Garc{\'\i}a}}, \ and\ \bibinfo {author} {\bibfnamefont
  {M.}~\bibnamefont {Taravillo}},\ }\bibfield  {title} {\enquote {\bibinfo
  {title} {Dft+ u calculations of crystal lattice, electronic structure, and
  phase stability under pressure of tio2 polymorphs},}\ }\href@noop {}
  {\bibfield  {journal} {\bibinfo  {journal} {The Journal of chemical physics}\
  }\textbf {\bibinfo {volume} {135}},\ \bibinfo {pages} {054503} (\bibinfo
  {year} {2011})}\BibitemShut {NoStop}%
\bibitem [{\citenamefont {Zhu}\ and\ \citenamefont {Gao}(2014)}]{Zhu2014}%
  \BibitemOpen
  \bibfield  {author} {\bibinfo {author} {\bibfnamefont {T.}~\bibnamefont
  {Zhu}}\ and\ \bibinfo {author} {\bibfnamefont {S.-P.}\ \bibnamefont {Gao}},\
  }\bibfield  {title} {\enquote {\bibinfo {title} {The stability, electronic
  structure, and optical property of TiO2 polymorphs},}\ }\href {\doibase
  10.1021/jp412462m} {\bibfield  {journal} {\bibinfo  {journal} {J.\ Phys.\
  Chem.~C}\ }\textbf {\bibinfo {volume} {118}},\ \bibinfo {pages}
  {11385--11396} (\bibinfo {year} {2014})}\BibitemShut {NoStop}%
\bibitem [{\citenamefont {Needs}\ \emph {et~al.}(2017)\citenamefont {Needs},
  \citenamefont {Lopez}, \citenamefont {Trail}, \citenamefont
  {Monserrat~Sanchez},\ and\ \citenamefont {Maezono}}]{Needs2017}%
  \BibitemOpen
  \bibfield  {author} {\bibinfo {author} {\bibfnamefont {R.}~\bibnamefont
  {Needs}}, \bibinfo {author} {\bibfnamefont {R.~P.}\ \bibnamefont {Lopez}},
  \bibinfo {author} {\bibfnamefont {J.}~\bibnamefont {Trail}}, \bibinfo
  {author} {\bibfnamefont {B.}~\bibnamefont {Monserrat~Sanchez}}, \ and\
  \bibinfo {author} {\bibfnamefont {R.}~\bibnamefont {Maezono}},\ }\bibfield
  {title} {\enquote {\bibinfo {title} {Research data supporting" quantum monte
  carlo study of the energetics of the rutile, anatase, brookite, and columbite
  tio2 polymorphs"},}\ }\href@noop {} {\  (\bibinfo {year} {2017})}\BibitemShut
  {NoStop}%
\bibitem [{\citenamefont {Matsui}\ and\ \citenamefont
  {Akaogi}(1991)}]{Matsui1991}%
  \BibitemOpen
  \bibfield  {author} {\bibinfo {author} {\bibfnamefont {M.}~\bibnamefont
  {Matsui}}\ and\ \bibinfo {author} {\bibfnamefont {M.}~\bibnamefont
  {Akaogi}},\ }\bibfield  {title} {\enquote {\bibinfo {title} {Molecular
  dynamics simulation of the structural and physical properties of the four
  polymorphs of {{\ce{TiO2}}}},}\ }\href {\doibase 10.1080/08927029108022432}
  {\bibfield  {journal} {\bibinfo  {journal} {Mol.\ Simul.}\ }\textbf {\bibinfo
  {volume} {6}},\ \bibinfo {pages} {239--244} (\bibinfo {year}
  {1991})}\BibitemShut {NoStop}%
\bibitem [{\citenamefont {Reinhardt}(2019)}]{Reinhardt2019}%
  \BibitemOpen
  \bibfield  {author} {\bibinfo {author} {\bibfnamefont {A.}~\bibnamefont
  {Reinhardt}},\ }\bibfield  {title} {\enquote {\bibinfo {title} {Phase
  behavior of empirical potentials of titanium dioxide},}\ }\href {\doibase
  10.1063/1.5115161} {\bibfield  {journal} {\bibinfo  {journal} {J.\ Chem.\
  Phys.}\ }\textbf {\bibinfo {volume} {151}},\ \bibinfo {pages} {064505}
  (\bibinfo {year} {2019})}\BibitemShut {NoStop}%
\bibitem [{\citenamefont {Reinhardt}, \citenamefont {Pickard},\ and\
  \citenamefont {Cheng}(2020)}]{Reinhardt2020}%
  \BibitemOpen
  \bibfield  {author} {\bibinfo {author} {\bibfnamefont {A.}~\bibnamefont
  {Reinhardt}}, \bibinfo {author} {\bibfnamefont {C.~J.}\ \bibnamefont
  {Pickard}}, \ and\ \bibinfo {author} {\bibfnamefont {B.}~\bibnamefont
  {Cheng}},\ }\bibfield  {title} {\enquote {\bibinfo {title} {Predicting the
  phase diagram of titanium dioxide with random search and pattern
  recognition},}\ }\href {\doibase 10.1039/D0CP02513E} {\bibfield  {journal}
  {\bibinfo  {journal} {Phys. Chem. Chem. Phys.}\ }\textbf {\bibinfo {volume}
  {22}},\ \bibinfo {pages} {12697--12705} (\bibinfo {year} {2020})}\BibitemShut
  {NoStop}%
\bibitem [{\citenamefont {Li}, \citenamefont {Xiao},\ and\ \citenamefont
  {Wang}(2018)}]{Li2018}%
  \BibitemOpen
  \bibfield  {author} {\bibinfo {author} {\bibfnamefont {Y.}~\bibnamefont
  {Li}}, \bibinfo {author} {\bibfnamefont {W.}~\bibnamefont {Xiao}}, \ and\
  \bibinfo {author} {\bibfnamefont {P.}~\bibnamefont {Wang}},\ }\bibfield
  {title} {{\enquote {\bibinfo {title}
  {Uncertainty quantification of artificial neural network based machine
  learning potentials},}\ }}in\ \href {\doibase 10.1115/IMECE2018-88071}
  {{\emph {\bibinfo {booktitle} {Materials}}}},\
  \bibinfo {series and number} {ASME International Mechanical Engineering
  Congress and Exposition, Proceedings (IMECE)}\ (\bibinfo  {publisher}
  {American Society of Mechanical Engineers (ASME)},\ \bibinfo {year} {2018})\
  \bibinfo {note} {aSME 2018 International Mechanical Engineering Congress and
  Exposition, IMECE 2018 ; Conference date: 09-11-2018 Through
  15-11-2018}\BibitemShut {NoStop}%
\bibitem [{\citenamefont {Ramakrishnan}\ \emph {et~al.}(2015)\citenamefont
  {Ramakrishnan}, \citenamefont {Dral}, \citenamefont {Rupp},\ and\
  \citenamefont {von Lilienfeld}}]{Ramakrishnan2015}%
  \BibitemOpen
  \bibfield  {author} {\bibinfo {author} {\bibfnamefont {R.}~\bibnamefont
  {Ramakrishnan}}, \bibinfo {author} {\bibfnamefont {P.~O.}\ \bibnamefont
  {Dral}}, \bibinfo {author} {\bibfnamefont {M.}~\bibnamefont {Rupp}}, \ and\
  \bibinfo {author} {\bibfnamefont {O.~A.}\ \bibnamefont {von Lilienfeld}},\
  }\bibfield  {title} {\enquote {\bibinfo {title} {Big data meets quantum
  chemistry approximations: The $\Delta$-machine learning approach},}\
  }\href {\doibase 10.1021/acs.jctc.5b00099} {\bibfield  {journal} {\bibinfo
  {journal} {Journal of Chemical Theory and Computation}\ }\textbf {\bibinfo
  {volume} {11}},\ \bibinfo {pages} {2087--2096} (\bibinfo {year}
  {2015})}\BibitemShut {NoStop}%
\bibitem [{\citenamefont {Perdew}\ \emph {et~al.}(2008)\citenamefont {Perdew},
  \citenamefont {Ruzsinszky}, \citenamefont {Csonka}, \citenamefont {Vydrov},
  \citenamefont {Scuseria}, \citenamefont {Constantin}, \citenamefont {Zhou},\
  and\ \citenamefont {Burke}}]{perdew2008restoring}%
  \BibitemOpen
  \bibfield  {author} {\bibinfo {author} {\bibfnamefont {J.~P.}\ \bibnamefont
  {Perdew}}, \bibinfo {author} {\bibfnamefont {A.}~\bibnamefont {Ruzsinszky}},
  \bibinfo {author} {\bibfnamefont {G.~I.}\ \bibnamefont {Csonka}}, \bibinfo
  {author} {\bibfnamefont {O.~A.}\ \bibnamefont {Vydrov}}, \bibinfo {author}
  {\bibfnamefont {G.~E.}\ \bibnamefont {Scuseria}}, \bibinfo {author}
  {\bibfnamefont {L.~A.}\ \bibnamefont {Constantin}}, \bibinfo {author}
  {\bibfnamefont {X.}~\bibnamefont {Zhou}}, \ and\ \bibinfo {author}
  {\bibfnamefont {K.}~\bibnamefont {Burke}},\ }\bibfield  {title} {\enquote
  {\bibinfo {title} {Restoring the density-gradient expansion for exchange in
  solids and surfaces},}\ }\href {\doibase 10.1103/PhysRevLett.100.136406}
  {\bibfield  {journal} {\bibinfo  {journal} {Phys.\ Rev.\ Lett.}\ }\textbf
  {\bibinfo {volume} {100}},\ \bibinfo {pages} {136406} (\bibinfo {year}
  {2008})}\BibitemShut {NoStop}%
\bibitem [{\citenamefont {Ding}\ and\ \citenamefont {Xiao}(2014)}]{Ding2014}%
  \BibitemOpen
  \bibfield  {author} {\bibinfo {author} {\bibfnamefont {Y.}~\bibnamefont
  {Ding}}\ and\ \bibinfo {author} {\bibfnamefont {B.}~\bibnamefont {Xiao}},\
  }\bibfield  {title} {\enquote {\bibinfo {title} {Anisotropic elasticity,
  sound velocity and thermal conductivity of tio2 polymorphs from first
  principles calculations},}\ }\href@noop {} {\bibfield  {journal} {\bibinfo
  {journal} {Computational materials science}\ }\textbf {\bibinfo {volume}
  {82}},\ \bibinfo {pages} {202--218} (\bibinfo {year} {2014})}\BibitemShut
  {NoStop}%
\bibitem [{\citenamefont {Lyle}, \citenamefont {Pickard},\ and\ \citenamefont
  {Needs}(2015)}]{lyle2015prediction}%
  \BibitemOpen
  \bibfield  {author} {\bibinfo {author} {\bibfnamefont {M.~J.}\ \bibnamefont
  {Lyle}}, \bibinfo {author} {\bibfnamefont {C.~J.}\ \bibnamefont {Pickard}}, \
  and\ \bibinfo {author} {\bibfnamefont {R.~J.}\ \bibnamefont {Needs}},\
  }\bibfield  {title} {\enquote {\bibinfo {title} {Prediction of 10-fold
  coordinated \ce{TiO2} and \ce{SiO2} structures at multimegabar pressures},}\
  }\href {\doibase 10.1073/pnas.1500604112} {\bibfield  {journal} {\bibinfo
  {journal} {Proc.\ Natl Acad.\ Sci.\ U.~S.~A.}\ }\textbf {\bibinfo {volume}
  {112}},\ \bibinfo {pages} {6898--6901} (\bibinfo {year} {2015})}\BibitemShut
  {NoStop}%
\bibitem [{\citenamefont {Trail}\ \emph {et~al.}(2017)\citenamefont {Trail},
  \citenamefont {Monserrat}, \citenamefont {R{\'\i}os}, \citenamefont
  {Maezono},\ and\ \citenamefont {Needs}}]{trail2017quantum}%
  \BibitemOpen
  \bibfield  {author} {\bibinfo {author} {\bibfnamefont {J.}~\bibnamefont
  {Trail}}, \bibinfo {author} {\bibfnamefont {B.}~\bibnamefont {Monserrat}},
  \bibinfo {author} {\bibfnamefont {P.~L.}\ \bibnamefont {R{\'\i}os}}, \bibinfo
  {author} {\bibfnamefont {R.}~\bibnamefont {Maezono}}, \ and\ \bibinfo
  {author} {\bibfnamefont {R.~J.}\ \bibnamefont {Needs}},\ }\bibfield  {title}
  {\enquote {\bibinfo {title} {Quantum monte carlo study of the energetics of
  the rutile, anatase, brookite, and columbite \ce{TiO2} polymorphs},}\ }\href
  {\doibase 10.1103/PhysRevB.95.121108} {\bibfield  {journal} {\bibinfo
  {journal} {Phys.\ Rev.\ B}\ }\textbf {\bibinfo {volume} {95}},\ \bibinfo
  {pages} {121108} (\bibinfo {year} {2017})}\BibitemShut {NoStop}%
\bibitem [{\citenamefont {Clark}\ \emph {et~al.}(2005)\citenamefont {Clark},
  \citenamefont {Segall}, \citenamefont {Pickard}, \citenamefont {Hasnip},
  \citenamefont {Probert}, \citenamefont {Refson},\ and\ \citenamefont
  {Payne}}]{clark2005first}%
  \BibitemOpen
  \bibfield  {author} {\bibinfo {author} {\bibfnamefont {S.~J.}\ \bibnamefont
  {Clark}}, \bibinfo {author} {\bibfnamefont {M.~D.}\ \bibnamefont {Segall}},
  \bibinfo {author} {\bibfnamefont {C.~J.}\ \bibnamefont {Pickard}}, \bibinfo
  {author} {\bibfnamefont {P.~J.}\ \bibnamefont {Hasnip}}, \bibinfo {author}
  {\bibfnamefont {M.~I.}\ \bibnamefont {Probert}}, \bibinfo {author}
  {\bibfnamefont {K.}~\bibnamefont {Refson}}, \ and\ \bibinfo {author}
  {\bibfnamefont {M.~C.}\ \bibnamefont {Payne}},\ }\bibfield  {title} {\enquote
  {\bibinfo {title} {First principles methods using {CASTEP}},}\ }\href
  {\doibase 10.1524/zkri.220.5.567.65075} {\bibfield  {journal} {\bibinfo
  {journal} {Z.\ Kristallogr.\ Cryst.\ Mater.}\ }\textbf {\bibinfo {volume}
  {220}},\ \bibinfo {pages} {567--570} (\bibinfo {year} {2005})}\BibitemShut
  {NoStop}%
\bibitem [{\citenamefont {Dubrovinskaia}\ \emph {et~al.}(2001)\citenamefont
  {Dubrovinskaia}, \citenamefont {Dubrovinsky}, \citenamefont {Ahuja},
  \citenamefont {Prokopenko}, \citenamefont {Dmitriev}, \citenamefont {Weber},
  \citenamefont {Osorio-Guillen},\ and\ \citenamefont
  {Johansson}}]{Dubrovinskaia2001}%
  \BibitemOpen
  \bibfield  {author} {\bibinfo {author} {\bibfnamefont {N.~A.}\ \bibnamefont
  {Dubrovinskaia}}, \bibinfo {author} {\bibfnamefont {L.~S.}\ \bibnamefont
  {Dubrovinsky}}, \bibinfo {author} {\bibfnamefont {R.}~\bibnamefont {Ahuja}},
  \bibinfo {author} {\bibfnamefont {V.~B.}\ \bibnamefont {Prokopenko}},
  \bibinfo {author} {\bibfnamefont {V.}~\bibnamefont {Dmitriev}}, \bibinfo
  {author} {\bibfnamefont {H.-P.}\ \bibnamefont {Weber}}, \bibinfo {author}
  {\bibfnamefont {J.~M.}\ \bibnamefont {Osorio-Guillen}}, \ and\ \bibinfo
  {author} {\bibfnamefont {B.}~\bibnamefont {Johansson}},\ }\bibfield  {title}
  {\enquote {\bibinfo {title} {Experimental and theoretical identification of a
  new high-pressure \ce{TiO2} polymorph},}\ }\href {\doibase
  10.1103/PhysRevLett.87.275501} {\bibfield  {journal} {\bibinfo  {journal}
  {Phys.\ Rev.\ Lett.}\ }\textbf {\bibinfo {volume} {87}},\ \bibinfo {pages}
  {275501} (\bibinfo {year} {2001})}\BibitemShut {NoStop}%
\bibitem [{\citenamefont {Arlt}\ \emph {et~al.}(2000)\citenamefont {Arlt},
  \citenamefont {Bermejo}, \citenamefont {Blanco}, \citenamefont {Gerward},
  \citenamefont {Jiang}, \citenamefont {Olsen},\ and\ \citenamefont
  {Recio}}]{Arlt2000}%
  \BibitemOpen
  \bibfield  {author} {\bibinfo {author} {\bibfnamefont {T.}~\bibnamefont
  {Arlt}}, \bibinfo {author} {\bibfnamefont {M.}~\bibnamefont {Bermejo}},
  \bibinfo {author} {\bibfnamefont {M.}~\bibnamefont {Blanco}}, \bibinfo
  {author} {\bibfnamefont {L.}~\bibnamefont {Gerward}}, \bibinfo {author}
  {\bibfnamefont {J.}~\bibnamefont {Jiang}}, \bibinfo {author} {\bibfnamefont
  {J.~S.}\ \bibnamefont {Olsen}}, \ and\ \bibinfo {author} {\bibfnamefont
  {J.}~\bibnamefont {Recio}},\ }\bibfield  {title} {\enquote {\bibinfo {title}
  {High-pressure polymorphs of anatase tio 2},}\ }\href@noop {} {\bibfield
  {journal} {\bibinfo  {journal} {Physical Review B}\ }\textbf {\bibinfo
  {volume} {61}},\ \bibinfo {pages} {14414} (\bibinfo {year}
  {2000})}\BibitemShut {NoStop}%
\bibitem [{\citenamefont {Mammone}, \citenamefont {Sharma},\ and\ \citenamefont
  {Nicol}(1980)}]{Mammone1980}%
  \BibitemOpen
  \bibfield  {author} {\bibinfo {author} {\bibfnamefont {J.}~\bibnamefont
  {Mammone}}, \bibinfo {author} {\bibfnamefont {S.}~\bibnamefont {Sharma}}, \
  and\ \bibinfo {author} {\bibfnamefont {M.}~\bibnamefont {Nicol}},\ }\bibfield
   {title} {\enquote {\bibinfo {title} {Raman study of rutile (tio2) at high
  pressures},}\ }\href@noop {} {\bibfield  {journal} {\bibinfo  {journal}
  {Solid State Communications}\ }\textbf {\bibinfo {volume} {34}},\ \bibinfo
  {pages} {799--802} (\bibinfo {year} {1980})}\BibitemShut {NoStop}%
\bibitem [{\citenamefont {Arashi}(1992)}]{Arashi1992}%
  \BibitemOpen
  \bibfield  {author} {\bibinfo {author} {\bibfnamefont {H.}~\bibnamefont
  {Arashi}},\ }\bibfield  {title} {\enquote {\bibinfo {title} {Raman
  spectroscopic study of the pressure-induced phase transition in tio2},}\
  }\href@noop {} {\bibfield  {journal} {\bibinfo  {journal} {Journal of Physics
  and Chemistry of Solids}\ }\textbf {\bibinfo {volume} {53}},\ \bibinfo
  {pages} {355--359} (\bibinfo {year} {1992})}\BibitemShut {NoStop}%
\bibitem [{\citenamefont {Gerward}\ and\ \citenamefont
  {Staun~Olsen}(1997)}]{Gerward1997}%
  \BibitemOpen
  \bibfield  {author} {\bibinfo {author} {\bibfnamefont {L.}~\bibnamefont
  {Gerward}}\ and\ \bibinfo {author} {\bibfnamefont {J.}~\bibnamefont
  {Staun~Olsen}},\ }\bibfield  {title} {\enquote {\bibinfo {title} {Post-rutile
  high-pressure phases in tio2},}\ }\href@noop {} {\bibfield  {journal}
  {\bibinfo  {journal} {Journal of Applied Crystallography}\ }\textbf {\bibinfo
  {volume} {30}},\ \bibinfo {pages} {259--264} (\bibinfo {year}
  {1997})}\BibitemShut {NoStop}%
\bibitem [{\citenamefont {Buckingham}(1938)}]{Buckingham1938}%
  \BibitemOpen
  \bibfield  {author} {\bibinfo {author} {\bibfnamefont {R.~A.}\ \bibnamefont
  {Buckingham}},\ }\bibfield  {title} {\enquote {\bibinfo {title} {The
  classical equation of state of gaseous helium, neon and argon},}\ }\href
  {\doibase 10.1098/rspa.1938.0173} {\bibfield  {journal} {\bibinfo  {journal}
  {Proc.\ R.\ Soc.\ Lond.~A}\ }\textbf {\bibinfo {volume} {168}},\ \bibinfo
  {pages} {264--283} (\bibinfo {year} {1938})}\BibitemShut {NoStop}%
\bibitem [{\citenamefont {Naicker}\ \emph {et~al.}(2005)\citenamefont
  {Naicker}, \citenamefont {Cummings}, \citenamefont {Zhang},\ and\
  \citenamefont {Banfield}}]{Naicker2005}%
  \BibitemOpen
  \bibfield  {author} {\bibinfo {author} {\bibfnamefont {P.~K.}\ \bibnamefont
  {Naicker}}, \bibinfo {author} {\bibfnamefont {P.~T.}\ \bibnamefont
  {Cummings}}, \bibinfo {author} {\bibfnamefont {H.}~\bibnamefont {Zhang}}, \
  and\ \bibinfo {author} {\bibfnamefont {J.~F.}\ \bibnamefont {Banfield}},\
  }\bibfield  {title} {\enquote {\bibinfo {title} {Characterization of titanium
  dioxide nanoparticles using molecular dynamics simulations},}\ }\href
  {\doibase 10.1021/jp050963q} {\bibfield  {journal} {\bibinfo  {journal} {J.\
  Phys.\ Chem.~B}\ }\textbf {\bibinfo {volume} {109}},\ \bibinfo {pages}
  {15243--15249} (\bibinfo {year} {2005})}\BibitemShut {NoStop}%
\bibitem [{\citenamefont {Koparde}\ and\ \citenamefont
  {Cummings}(2005)}]{Koparde2005}%
  \BibitemOpen
  \bibfield  {author} {\bibinfo {author} {\bibfnamefont {V.~N.}\ \bibnamefont
  {Koparde}}\ and\ \bibinfo {author} {\bibfnamefont {P.~T.}\ \bibnamefont
  {Cummings}},\ }\bibfield  {title} {\enquote {\bibinfo {title} {Molecular
  dynamics simulation of titanium dioxide nanoparticle sintering},}\ }\href
  {\doibase 10.1021/jp054667p} {\bibfield  {journal} {\bibinfo  {journal} {The
  Journal of Physical Chemistry B}\ }\textbf {\bibinfo {volume} {109}},\
  \bibinfo {pages} {24280--24287} (\bibinfo {year} {2005})},\ \bibinfo {note}
  {pMID: 16375425},\ \Eprint
  {http://arxiv.org/abs/https://doi.org/10.1021/jp054667p}
  {https://doi.org/10.1021/jp054667p} \BibitemShut {NoStop}%
\bibitem [{\citenamefont {Heyhat}, \citenamefont {Abbasi},\ and\ \citenamefont
  {Rajabpour}(2021)}]{Heyhat2021}%
  \BibitemOpen
  \bibfield  {author} {\bibinfo {author} {\bibfnamefont {M.}~\bibnamefont
  {Heyhat}}, \bibinfo {author} {\bibfnamefont {M.}~\bibnamefont {Abbasi}}, \
  and\ \bibinfo {author} {\bibfnamefont {A.}~\bibnamefont {Rajabpour}},\
  }\bibfield  {title} {\enquote {\bibinfo {title} {Molecular dynamic simulation
  on the density of titanium dioxide and silver water-based nanofluids using
  ternary mixture model},}\ }\href {\doibase
  https://doi.org/10.1016/j.molliq.2021.115966} {\bibfield  {journal} {\bibinfo
   {journal} {Journal of Molecular Liquids}\ }\textbf {\bibinfo {volume}
  {333}},\ \bibinfo {pages} {115966} (\bibinfo {year} {2021})}\BibitemShut
  {NoStop}%
\bibitem [{\citenamefont {Xu}\ \emph {et~al.}(2016)\citenamefont {Xu},
  \citenamefont {Wang}, \citenamefont {Hu}, \citenamefont {Bell},\ and\
  \citenamefont {Yan}}]{Xu2016}%
  \BibitemOpen
  \bibfield  {author} {\bibinfo {author} {\bibfnamefont {Y.}~\bibnamefont
  {Xu}}, \bibinfo {author} {\bibfnamefont {M.}~\bibnamefont {Wang}}, \bibinfo
  {author} {\bibfnamefont {N.}~\bibnamefont {Hu}}, \bibinfo {author}
  {\bibfnamefont {J.}~\bibnamefont {Bell}}, \ and\ \bibinfo {author}
  {\bibfnamefont {C.}~\bibnamefont {Yan}},\ }\bibfield  {title} {\enquote
  {\bibinfo {title} {Atomistic investigation into the mechanical behaviour of
  crystalline and amorphous tio 2 nanotubes},}\ }\href@noop {} {\bibfield
  {journal} {\bibinfo  {journal} {RSC advances}\ }\textbf {\bibinfo {volume}
  {6}},\ \bibinfo {pages} {28121--28129} (\bibinfo {year} {2016})}\BibitemShut
  {NoStop}%
\bibitem [{\citenamefont {Behler}\ and\ \citenamefont
  {Parrinello}(2007)}]{Behler2007}%
  \BibitemOpen
  \bibfield  {author} {\bibinfo {author} {\bibfnamefont {J.}~\bibnamefont
  {Behler}}\ and\ \bibinfo {author} {\bibfnamefont {M.}~\bibnamefont
  {Parrinello}},\ }\bibfield  {title} {\enquote {\bibinfo {title} {Generalized
  neural-network representation of high-dimensional potential-energy
  surfaces},}\ }\href {\doibase 10.1103/PhysRevLett.98.146401} {\bibfield
  {journal} {\bibinfo  {journal} {Phys. Rev. Lett.}\ }\textbf {\bibinfo
  {volume} {98}},\ \bibinfo {pages} {146401} (\bibinfo {year}
  {2007})}\BibitemShut {NoStop}%
\bibitem [{\citenamefont {Cheng}\ \emph {et~al.}(2020)\citenamefont {Cheng},
  \citenamefont {Griffiths}, \citenamefont {Wengert}, \citenamefont {Kunkel},
  \citenamefont {Stenczel}, \citenamefont {Zhu}, \citenamefont {Deringer},
  \citenamefont {Bernstein}, \citenamefont {Margraf}, \citenamefont {Reuter},\
  and\ \citenamefont {Csanyi}}]{Cheng2020asap}%
  \BibitemOpen
  \bibfield  {author} {\bibinfo {author} {\bibfnamefont {B.}~\bibnamefont
  {Cheng}}, \bibinfo {author} {\bibfnamefont {R.-R.}\ \bibnamefont
  {Griffiths}}, \bibinfo {author} {\bibfnamefont {S.}~\bibnamefont {Wengert}},
  \bibinfo {author} {\bibfnamefont {C.}~\bibnamefont {Kunkel}}, \bibinfo
  {author} {\bibfnamefont {T.}~\bibnamefont {Stenczel}}, \bibinfo {author}
  {\bibfnamefont {B.}~\bibnamefont {Zhu}}, \bibinfo {author} {\bibfnamefont
  {V.~L.}\ \bibnamefont {Deringer}}, \bibinfo {author} {\bibfnamefont
  {N.}~\bibnamefont {Bernstein}}, \bibinfo {author} {\bibfnamefont {J.~T.}\
  \bibnamefont {Margraf}}, \bibinfo {author} {\bibfnamefont {K.}~\bibnamefont
  {Reuter}}, \ and\ \bibinfo {author} {\bibfnamefont {G.}~\bibnamefont
  {Csanyi}},\ }\bibfield  {title} {\enquote {\bibinfo {title} {Mapping
  materials and molecules},}\ }\href {\doibase 10.1021/acs.accounts.0c00403}
  {\bibfield  {journal} {\bibinfo  {journal} {Accounts of Chemical Research}\
  }\textbf {\bibinfo {volume} {53}},\ \bibinfo {pages} {1981--1991} (\bibinfo
  {year} {2020})},\ \bibinfo {note} {pMID: 32794697},\ \Eprint
  {http://arxiv.org/abs/https://doi.org/10.1021/acs.accounts.0c00403}
  {https://doi.org/10.1021/acs.accounts.0c00403} \BibitemShut {NoStop}%
\bibitem [{\citenamefont {Cersonsky}\ \emph {et~al.}(2021)\citenamefont
  {Cersonsky}, \citenamefont {Helfrecht}, \citenamefont {Engel}, \citenamefont
  {Kliavinek},\ and\ \citenamefont {Ceriotti}}]{Cersonsky2021}%
  \BibitemOpen
  \bibfield  {author} {\bibinfo {author} {\bibfnamefont {R.~K.}\ \bibnamefont
  {Cersonsky}}, \bibinfo {author} {\bibfnamefont {B.}~\bibnamefont
  {Helfrecht}}, \bibinfo {author} {\bibfnamefont {E.~A.}\ \bibnamefont
  {Engel}}, \bibinfo {author} {\bibfnamefont {S.}~\bibnamefont {Kliavinek}}, \
  and\ \bibinfo {author} {\bibfnamefont {M.}~\bibnamefont {Ceriotti}},\
  }\bibfield  {title} {\enquote {\bibinfo {title} {Improving sample and feature
  selection with principal covariates regression},}\ }\href@noop {} {\bibfield
  {journal} {\bibinfo  {journal} {Machine Learning: Science and Technology}\ }
  (\bibinfo {year} {2021})}\BibitemShut {NoStop}%
\bibitem [{\citenamefont {Imbalzano}\ \emph {et~al.}(2018)\citenamefont
  {Imbalzano}, \citenamefont {Anelli}, \citenamefont {Giofré}, \citenamefont
  {Klees}, \citenamefont {Behler},\ and\ \citenamefont
  {Ceriotti}}]{Imbalzano2018}%
  \BibitemOpen
  \bibfield  {author} {\bibinfo {author} {\bibfnamefont {G.}~\bibnamefont
  {Imbalzano}}, \bibinfo {author} {\bibfnamefont {A.}~\bibnamefont {Anelli}},
  \bibinfo {author} {\bibfnamefont {D.}~\bibnamefont {Giofré}}, \bibinfo
  {author} {\bibfnamefont {S.}~\bibnamefont {Klees}}, \bibinfo {author}
  {\bibfnamefont {J.}~\bibnamefont {Behler}}, \ and\ \bibinfo {author}
  {\bibfnamefont {M.}~\bibnamefont {Ceriotti}},\ }\bibfield  {title} {\enquote
  {\bibinfo {title} {Automatic selection of atomic fingerprints and reference
  configurations for machine-learning potentials},}\ }\href {\doibase
  10.1063/1.5024611} {\bibfield  {journal} {\bibinfo  {journal} {The Journal of
  Chemical Physics}\ }\textbf {\bibinfo {volume} {148}},\ \bibinfo {pages}
  {241730} (\bibinfo {year} {2018})},\ \Eprint
  {http://arxiv.org/abs/https://doi.org/10.1063/1.5024611}
  {https://doi.org/10.1063/1.5024611} \BibitemShut {NoStop}%
\bibitem [{\citenamefont {Bartók}\ and\ \citenamefont
  {Csányi}(2015)}]{Albert2015}%
  \BibitemOpen
  \bibfield  {author} {\bibinfo {author} {\bibfnamefont {A.~P.}\ \bibnamefont
  {Bartók}}\ and\ \bibinfo {author} {\bibfnamefont {G.}~\bibnamefont
  {Csányi}},\ }\bibfield  {title} {\enquote {\bibinfo {title} {Gaussian
  approximation potentials: A brief tutorial introduction},}\ }\href {\doibase
  https://doi.org/10.1002/qua.24927} {\bibfield  {journal} {\bibinfo  {journal}
  {International Journal of Quantum Chemistry}\ }\textbf {\bibinfo {volume}
  {115}},\ \bibinfo {pages} {1051--1057} (\bibinfo {year} {2015})},\ \Eprint
  {http://arxiv.org/abs/https://onlinelibrary.wiley.com/doi/pdf/10.1002/qua.24927}
  {https://onlinelibrary.wiley.com/doi/pdf/10.1002/qua.24927} \BibitemShut
  {NoStop}%
\bibitem [{\citenamefont {Plimpton}(1995)}]{Plimpton1995}%
  \BibitemOpen
  \bibfield  {author} {\bibinfo {author} {\bibfnamefont {S.}~\bibnamefont
  {Plimpton}},\ }\bibfield  {title} {\enquote {\bibinfo {title} {Fast parallel
  algorithms for short-range molecular dynamics},}\ }\href {\doibase
  10.1006/jcph.1995.1039} {\bibfield  {journal} {\bibinfo  {journal} {J.\
  Comput.\ Phys.}\ }\textbf {\bibinfo {volume} {117}},\ \bibinfo {pages}
  {1--19} (\bibinfo {year} {1995})}\BibitemShut {NoStop}%
\bibitem [{\citenamefont {Singraber}\ \emph {et~al.}(2019)\citenamefont
  {Singraber}, \citenamefont {Morawietz}, \citenamefont {Behler},\ and\
  \citenamefont {Dellago}}]{Singraber2019}%
  \BibitemOpen
  \bibfield  {author} {\bibinfo {author} {\bibfnamefont {A.}~\bibnamefont
  {Singraber}}, \bibinfo {author} {\bibfnamefont {T.}~\bibnamefont
  {Morawietz}}, \bibinfo {author} {\bibfnamefont {J.}~\bibnamefont {Behler}}, \
  and\ \bibinfo {author} {\bibfnamefont {C.}~\bibnamefont {Dellago}},\
  }\bibfield  {title} {\enquote {\bibinfo {title} {Parallel multistream
  training of high-dimensional neural network potentials},}\ }\href {\doibase
  10.1021/acs.jctc.8b01092} {\bibfield  {journal} {\bibinfo  {journal} {Journal
  of Chemical Theory and Computation}\ }\textbf {\bibinfo {volume} {15}},\
  \bibinfo {pages} {3075--3092} (\bibinfo {year} {2019})},\ \bibinfo {note}
  {pMID: 30995035},\ \Eprint
  {http://arxiv.org/abs/https://doi.org/10.1021/acs.jctc.8b01092}
  {https://doi.org/10.1021/acs.jctc.8b01092} \BibitemShut {NoStop}%
\bibitem [{\citenamefont {Cheng}\ and\ \citenamefont
  {Ceriotti}(2018)}]{Cheng2018}%
  \BibitemOpen
  \bibfield  {author} {\bibinfo {author} {\bibfnamefont {B.}~\bibnamefont
  {Cheng}}\ and\ \bibinfo {author} {\bibfnamefont {M.}~\bibnamefont
  {Ceriotti}},\ }\bibfield  {title} {\enquote {\bibinfo {title} {Computing the
  absolute gibbs free energy in atomistic simulations: Applications to defects
  in solids},}\ }\href {\doibase 10.1103/PhysRevB.97.054102} {\bibfield
  {journal} {\bibinfo  {journal} {Phys. Rev. B}\ }\textbf {\bibinfo {volume}
  {97}},\ \bibinfo {pages} {054102} (\bibinfo {year} {2018})}\BibitemShut
  {NoStop}%
\bibitem [{\citenamefont {Jamieson}\ and\ \citenamefont
  {Olinger}(1969)}]{Jamieson1969}%
  \BibitemOpen
  \bibfield  {author} {\bibinfo {author} {\bibfnamefont {J.~C.}\ \bibnamefont
  {Jamieson}}\ and\ \bibinfo {author} {\bibfnamefont {B.}~\bibnamefont
  {Olinger}},\ }\bibfield  {title} {\enquote {\bibinfo {title}
  {{Pressure-temperature studies of anatase, brookite rutile, and Ti02(II): A
  discussion}},}\ }\href@noop {} {\bibfield  {journal} {\bibinfo  {journal}
  {American Mineralogist}\ }\textbf {\bibinfo {volume} {54}},\ \bibinfo {pages}
  {1477--1481} (\bibinfo {year} {1969})},\ \Eprint
  {http://arxiv.org/abs/https://pubs.geoscienceworld.org/ammin/article-pdf/54/9-10/1477/4249594/am-1969-1477.pdf}
  {https://pubs.geoscienceworld.org/ammin/article-pdf/54/9-10/1477/4249594/am-1969-1477.pdf}
  \BibitemShut {NoStop}%
\bibitem [{\citenamefont {Che}\ \emph {et~al.}(2016)\citenamefont {Che},
  \citenamefont {Li}, \citenamefont {Zheng}, \citenamefont {Li},\ and\
  \citenamefont {Shi}}]{Che2016}%
  \BibitemOpen
  \bibfield  {author} {\bibinfo {author} {\bibfnamefont {X.}~\bibnamefont
  {Che}}, \bibinfo {author} {\bibfnamefont {L.}~\bibnamefont {Li}}, \bibinfo
  {author} {\bibfnamefont {J.}~\bibnamefont {Zheng}}, \bibinfo {author}
  {\bibfnamefont {G.}~\bibnamefont {Li}}, \ and\ \bibinfo {author}
  {\bibfnamefont {Q.}~\bibnamefont {Shi}},\ }\bibfield  {title} {\enquote
  {\bibinfo {title} {Heat capacity and thermodynamic functions of brookite
  tio2},}\ }\href {\doibase https://doi.org/10.1016/j.jct.2015.09.018}
  {\bibfield  {journal} {\bibinfo  {journal} {The Journal of Chemical
  Thermodynamics}\ }\textbf {\bibinfo {volume} {93}},\ \bibinfo {pages}
  {45--51} (\bibinfo {year} {2016})}\BibitemShut {NoStop}%
\bibitem [{\citenamefont {Swamy}, \citenamefont {Dubrovinskaia},\ and\
  \citenamefont {Dubrovinsky}(2002)}]{Swamy2002}%
  \BibitemOpen
  \bibfield  {author} {\bibinfo {author} {\bibfnamefont {V.}~\bibnamefont
  {Swamy}}, \bibinfo {author} {\bibfnamefont {N.~A.}\ \bibnamefont
  {Dubrovinskaia}}, \ and\ \bibinfo {author} {\bibfnamefont {L.~S.}\
  \bibnamefont {Dubrovinsky}},\ }\bibfield  {title} {\enquote {\bibinfo {title}
  {Compressibility of baddeleyite-type tio2 from static compression to 40
  gpa},}\ }\href {\doibase https://doi.org/10.1016/S0925-8388(02)00109-3}
  {\bibfield  {journal} {\bibinfo  {journal} {Journal of Alloys and Compounds}\
  }\textbf {\bibinfo {volume} {340}},\ \bibinfo {pages} {46--48} (\bibinfo
  {year} {2002})}\BibitemShut {NoStop}%
\bibitem [{\citenamefont {Swamy}\ \emph {et~al.}(2003)\citenamefont {Swamy},
  \citenamefont {Dubrovinsky}, \citenamefont {Dubrovinskaia}, \citenamefont
  {Simionovici}, \citenamefont {Drakopoulos}, \citenamefont {Dmitriev},\ and\
  \citenamefont {Weber}}]{Swamy2003}%
  \BibitemOpen
  \bibfield  {author} {\bibinfo {author} {\bibfnamefont {V.}~\bibnamefont
  {Swamy}}, \bibinfo {author} {\bibfnamefont {L.~S.}\ \bibnamefont
  {Dubrovinsky}}, \bibinfo {author} {\bibfnamefont {N.~A.}\ \bibnamefont
  {Dubrovinskaia}}, \bibinfo {author} {\bibfnamefont {A.~S.}\ \bibnamefont
  {Simionovici}}, \bibinfo {author} {\bibfnamefont {M.}~\bibnamefont
  {Drakopoulos}}, \bibinfo {author} {\bibfnamefont {V.}~\bibnamefont
  {Dmitriev}}, \ and\ \bibinfo {author} {\bibfnamefont {H.-P.}\ \bibnamefont
  {Weber}},\ }\bibfield  {title} {\enquote {\bibinfo {title} {Compression
  behavior of nanocrystalline anatase tio2},}\ }\href {\doibase
  https://doi.org/10.1016/S0038-1098(02)00601-4} {\bibfield  {journal}
  {\bibinfo  {journal} {Solid State Communications}\ }\textbf {\bibinfo
  {volume} {125}},\ \bibinfo {pages} {111--115} (\bibinfo {year}
  {2003})}\BibitemShut {NoStop}%
\bibitem [{\citenamefont {Deringer}\ \emph {et~al.}(2021)\citenamefont
  {Deringer}, \citenamefont {Bernstein}, \citenamefont {Cs{\'a}nyi},
  \citenamefont {Mahmoud}, \citenamefont {Ceriotti}, \citenamefont {Wilson},
  \citenamefont {Drabold},\ and\ \citenamefont {Elliott}}]{Deringer2021}%
  \BibitemOpen
  \bibfield  {author} {\bibinfo {author} {\bibfnamefont {V.~L.}\ \bibnamefont
  {Deringer}}, \bibinfo {author} {\bibfnamefont {N.}~\bibnamefont {Bernstein}},
  \bibinfo {author} {\bibfnamefont {G.}~\bibnamefont {Cs{\'a}nyi}}, \bibinfo
  {author} {\bibfnamefont {C.~B.}\ \bibnamefont {Mahmoud}}, \bibinfo {author}
  {\bibfnamefont {M.}~\bibnamefont {Ceriotti}}, \bibinfo {author}
  {\bibfnamefont {M.}~\bibnamefont {Wilson}}, \bibinfo {author} {\bibfnamefont
  {D.~A.}\ \bibnamefont {Drabold}}, \ and\ \bibinfo {author} {\bibfnamefont
  {S.~R.}\ \bibnamefont {Elliott}},\ }\bibfield  {title} {\enquote {\bibinfo
  {title} {Origins of structural and electronic transitions in disordered
  silicon},}\ }\href@noop {} {\bibfield  {journal} {\bibinfo  {journal}
  {Nature}\ }\textbf {\bibinfo {volume} {589}},\ \bibinfo {pages} {59--64}
  (\bibinfo {year} {2021})}\BibitemShut {NoStop}%
\bibitem [{\citenamefont {Teter}, \citenamefont {Payne},\ and\ \citenamefont
  {Allan}(1989)}]{Teter1989}%
  \BibitemOpen
  \bibfield  {author} {\bibinfo {author} {\bibfnamefont {M.~P.}\ \bibnamefont
  {Teter}}, \bibinfo {author} {\bibfnamefont {M.~C.}\ \bibnamefont {Payne}}, \
  and\ \bibinfo {author} {\bibfnamefont {D.~C.}\ \bibnamefont {Allan}},\
  }\bibfield  {title} {\enquote {\bibinfo {title} {Solution of schr\"odinger's
  equation for large systems},}\ }\href {\doibase 10.1103/PhysRevB.40.12255}
  {\bibfield  {journal} {\bibinfo  {journal} {Phys. Rev. B}\ }\textbf {\bibinfo
  {volume} {40}},\ \bibinfo {pages} {12255--12263} (\bibinfo {year}
  {1989})}\BibitemShut {NoStop}%
\bibitem [{\citenamefont {Bussi}, \citenamefont {Donadio},\ and\ \citenamefont
  {Parrinello}(2007)}]{Bussi2007}%
  \BibitemOpen
  \bibfield  {author} {\bibinfo {author} {\bibfnamefont {G.}~\bibnamefont
  {Bussi}}, \bibinfo {author} {\bibfnamefont {D.}~\bibnamefont {Donadio}}, \
  and\ \bibinfo {author} {\bibfnamefont {M.}~\bibnamefont {Parrinello}},\
  }\bibfield  {title} {\enquote {\bibinfo {title} {Canonical sampling through
  velocity rescaling},}\ }\href {\doibase 10.1063/1.2408420} {\bibfield
  {journal} {\bibinfo  {journal} {J.\ Chem.\ Phys.}\ }\textbf {\bibinfo
  {volume} {126}},\ \bibinfo {pages} {014101} (\bibinfo {year}
  {2007})}\BibitemShut {NoStop}%
\bibitem [{\citenamefont {Kapil}\ \emph {et~al.}(2019)\citenamefont {Kapil},
  \citenamefont {Rossi}, \citenamefont {Marsalek}, \citenamefont {Petraglia},
  \citenamefont {Litman}, \citenamefont {Spura}, \citenamefont {Cheng},
  \citenamefont {Cuzzocrea}, \citenamefont {Meißner}, \citenamefont {Wilkins},
  \citenamefont {Helfrecht}, \citenamefont {Juda}, \citenamefont {Bienvenue},
  \citenamefont {Fang}, \citenamefont {Kessler}, \citenamefont {Poltavsky},
  \citenamefont {Vandenbrande}, \citenamefont {Wieme}, \citenamefont
  {Corminboeuf}, \citenamefont {Kühne}, \citenamefont {Manolopoulos},
  \citenamefont {Markland}, \citenamefont {Richardson}, \citenamefont
  {Tkatchenko}, \citenamefont {Tribello}, \citenamefont {{Van Speybroeck}},\
  and\ \citenamefont {Ceriotti}}]{Kapil2019}%
  \BibitemOpen
  \bibfield  {author} {\bibinfo {author} {\bibfnamefont {V.}~\bibnamefont
  {Kapil}}, \bibinfo {author} {\bibfnamefont {M.}~\bibnamefont {Rossi}},
  \bibinfo {author} {\bibfnamefont {O.}~\bibnamefont {Marsalek}}, \bibinfo
  {author} {\bibfnamefont {R.}~\bibnamefont {Petraglia}}, \bibinfo {author}
  {\bibfnamefont {Y.}~\bibnamefont {Litman}}, \bibinfo {author} {\bibfnamefont
  {T.}~\bibnamefont {Spura}}, \bibinfo {author} {\bibfnamefont
  {B.}~\bibnamefont {Cheng}}, \bibinfo {author} {\bibfnamefont
  {A.}~\bibnamefont {Cuzzocrea}}, \bibinfo {author} {\bibfnamefont {R.~H.}\
  \bibnamefont {Meißner}}, \bibinfo {author} {\bibfnamefont {D.~M.}\
  \bibnamefont {Wilkins}}, \bibinfo {author} {\bibfnamefont {B.~A.}\
  \bibnamefont {Helfrecht}}, \bibinfo {author} {\bibfnamefont {P.}~\bibnamefont
  {Juda}}, \bibinfo {author} {\bibfnamefont {S.~P.}\ \bibnamefont {Bienvenue}},
  \bibinfo {author} {\bibfnamefont {W.}~\bibnamefont {Fang}}, \bibinfo {author}
  {\bibfnamefont {J.}~\bibnamefont {Kessler}}, \bibinfo {author} {\bibfnamefont
  {I.}~\bibnamefont {Poltavsky}}, \bibinfo {author} {\bibfnamefont
  {S.}~\bibnamefont {Vandenbrande}}, \bibinfo {author} {\bibfnamefont
  {J.}~\bibnamefont {Wieme}}, \bibinfo {author} {\bibfnamefont
  {C.}~\bibnamefont {Corminboeuf}}, \bibinfo {author} {\bibfnamefont {T.~D.}\
  \bibnamefont {Kühne}}, \bibinfo {author} {\bibfnamefont {D.~E.}\
  \bibnamefont {Manolopoulos}}, \bibinfo {author} {\bibfnamefont {T.~E.}\
  \bibnamefont {Markland}}, \bibinfo {author} {\bibfnamefont {J.~O.}\
  \bibnamefont {Richardson}}, \bibinfo {author} {\bibfnamefont
  {A.}~\bibnamefont {Tkatchenko}}, \bibinfo {author} {\bibfnamefont {G.~A.}\
  \bibnamefont {Tribello}}, \bibinfo {author} {\bibfnamefont {V.}~\bibnamefont
  {{Van Speybroeck}}}, \ and\ \bibinfo {author} {\bibfnamefont
  {M.}~\bibnamefont {Ceriotti}},\ }\bibfield  {title} {\enquote {\bibinfo
  {title} {i-pi 2.0: A universal force engine for advanced molecular
  simulations},}\ }\href {\doibase https://doi.org/10.1016/j.cpc.2018.09.020}
  {\bibfield  {journal} {\bibinfo  {journal} {Computer Physics Communications}\
  }\textbf {\bibinfo {volume} {236}},\ \bibinfo {pages} {214--223} (\bibinfo
  {year} {2019})}\BibitemShut {NoStop}%
\bibitem [{\citenamefont {Ceriotti}, \citenamefont {More},\ and\ \citenamefont
  {Manolopoulos}(2014)}]{Ceriotti2014}%
  \BibitemOpen
  \bibfield  {author} {\bibinfo {author} {\bibfnamefont {M.}~\bibnamefont
  {Ceriotti}}, \bibinfo {author} {\bibfnamefont {J.}~\bibnamefont {More}}, \
  and\ \bibinfo {author} {\bibfnamefont {D.~E.}\ \bibnamefont {Manolopoulos}},\
  }\bibfield  {title} {\enquote {\bibinfo {title} {i-pi: A python interface for
  ab initio path integral molecular dynamics simulations},}\ }\href {\doibase
  https://doi.org/10.1016/j.cpc.2013.10.027} {\bibfield  {journal} {\bibinfo
  {journal} {Computer Physics Communications}\ }\textbf {\bibinfo {volume}
  {185}},\ \bibinfo {pages} {1019--1026} (\bibinfo {year} {2014})}\BibitemShut
  {NoStop}%
\end{thebibliography}
\end{document}